\documentclass[review,times]{elsarticle}
\journal{Computer Physics Communications}

\usepackage{moreverb,url}
\usepackage{amsmath}
\usepackage{booktabs}
\usepackage{tabularx}

\usepackage[colorlinks,bookmarksopen,bookmarksnumbered]{hyperref}
\usepackage{xco lor}
\usepackage{subcaption}
\usepackage{listings}
\usepackage{xr}
\begin{document}

\begin{frontmatter}

\title{Code modernization strategies for short-range non-bonded molecular dynamics simulations}

\author[affil1]{James Vance\corref{mycorrespondingauthor}}
\cortext[mycorrespondingauthor]{Corresponding author}
\ead{jnvance@outlook.com}

\author[affil1]{Zhen-Hao Xu}
\author[affil1]{Nikita Tretyakov}
\author[affil2]{Torsten Stuehn}
\author[affil3]{Markus Rampp}
\author[affil3]{Sebastian Eibl}
\author[affil4]{Christoph Junghans}
\author[affil1]{Andr\'e Brinkmann}

\address[affil1]{Zentrum f\"ur Datenverarbeitung, Johannes Gutenberg-Universit\"at Mainz, Mainz, Germany}
\address[affil2]{Max Planck Institute for Polymer Research, Mainz, Germany}
\address[affil3]{Max Planck Computing and Data Facility, Garching, Germany}
\address[affil4]{Applied Computer Science Group, Los Alamos National Laboratory, Los Alamos, New Mexico, USA}

\begin{abstract}
Modern HPC systems are increasingly relying on greater core counts and wider vector registers. Thus, applications need to be adapted to fully utilize these hardware capabilities.
One class of applications that can benefit from this increase in parallelism are molecular dynamics simulations.
In this paper, we describe our efforts at modernizing the ESPResSo++ simulation package for molecular dynamics by restructuring its particle data layout for efficient memory accesses and applying vectorization techniques to benefit the calculation of short-range non-bonded forces, which results in an overall three times speedup and serves as a baseline for further optimizations.
We also implement fine-grained parallelism for multi-core CPUs through HPX, a C++ runtime system which uses lightweight threads and an asynchronous many-task approach to maximize concurrency.
Our goal is to evaluate the performance of an HPX-based approach compared to the bulk-synchronous MPI-based implementation.
This requires the introduction of an additional layer to the domain decomposition scheme that defines the task granularity.
On spatially inhomogeneous systems, which impose a corresponding load-imbalance in traditional MPI-based approaches, we demonstrate that by choosing an optimal task size, the efficient work-stealing mechanisms of HPX can overcome the overhead of communication resulting in an overall 1.4 times speedup compared to the baseline MPI version.
\end{abstract}


\end{frontmatter}


\section{Introduction}

As the growth of processor frequency continues to plateau, modern HPC systems increasingly rely on greater concurrency and parallelism to deliver more performance.
This comes in the form of increasing core counts and providing wider vector registers.
However, as core counts increase, traditional parallelization methods that rely on MPI contend with fewer available memory per core and their performance faces increased sensitivity to load imbalances and synchronization mechanisms.
Full utilization of wider vector registers also means critical parts of applications need to be rewritten and often requires more complex solutions.
Thus, applications have to be adapted and modernized in order to fully maximize new hardware capabilities and overcome memory constraints.

Molecular dynamics (MD) simulations play a significant role in the study and discovery of new materials especially in soft matter research.
They also act as an ideal example to harness more recent hardware capabilities since the calculations involved offer different ways of parallelization, such as through force decomposition, atom decomposition and spatial decomposition schemes \citep{plimpton1995fast}.
Consequently, many simulation packages have been developed over the past decades that can run on machines with up to hundreds of thousands of cores such as LAMMPS \citep{thompson2022lammps}, GROMACS \citep{abraham2015gromacs}, and NAMD~\citep{phillips2020scalable}.

In this paper, we focus on ESPResSo++, an open-source software for performing molecular dynamics simulations of condensed soft matter systems \citep{halverson2013espressopp, guzman2019espresso++}.
It is built on a C++ backend with MPI-based communication enabling fast parallel execution.
It also provides a Python frontend for convenient scripting and analysis.
The foremost guideline in the design of ESPResSo++ is extensibility which allowed it to become a sandbox in which users can easily develop new methods and algorithms \citep{halverson2013espressopp}.
However, this means that decisions taken in favor of extensibility may not always result in the best optimized code for performance on modern hardware.

To address this, we apply SIMD vectorization and multithreaded task-based parallelism to improve the performance of ESPResSo++ on modern multi-core CPUs.
We specifically focus on molecular dynamics simulations involving short-range non-bonded forces whose calculations take up a significant portion of the total simulation time and can benefit the most from these additional layers of parallelism.

First, we maximize the utilization of wider vector registers through SIMD vectorization.
SIMD, which stands for single instruction multiple data, allows a single instruction to simultaneously process multiple vector elements whose size depends on the bit length of registers known as the SIMD width \citep{watanabe2019simd}.
For example, Intel has been developing extensions of the x86\_64 instruction set over the years including Streaming SIMD Extensions (SSE) which support 128-bit registers, Advanced Vector Extensions~2 (AVX) for 256-bit registers, and AVX-512 for 512-bit registers.
These are often provided as low-level functions called intrinsics which can be invoked directly from the source code but whose availability may vary among different architectures\footnote{\url{https://software.intel.com/sites/landingpage/IntrinsicsGuide}}.
To benefit from these extensions while retaining code portability, other approaches can be used such as directives that give hints to the compiler about vectorizable loops such as \texttt{\#pragma vector} provided by Intel\footnote{\url{https://software.intel.com/content/dam/develop/external/us/en/documents/cpp\_compiler\_classic.pdf}} and \texttt{\#pragma omp simd} from OpenMP\footnote{\url{https://www.openmp.org/spec-html/5.1/openmp.html}}, and libraries that implement generalized high-level abstractions to the underlying intrinsics such as the Vc library \citep{kretz2012vc}. SIMD optimizations can improve the performance of MD simulations independent of the application domain and we have evaluated their impact based on a standard Lennard-Jones fluid simulation and a polymer melt simulation running on a single node as benchmarks.  
 
Next, we  address the need for thread-level parallelism.
Many applications running on computing clusters, including ESPResSo++, rely on the MPI programming paradigm that explicitly synchronizes calculation steps through communication.
However, as the number of cores per node increases, applications become more sensitive to load imbalances and operating system noise causing wait times to increase.
Such scenarios prevent good application scaling for inhomogeneous systems unless a good load-balancing strategy is employed~\citep{ishiyama2012fflops,fattebert2012dynamic, Eibl2019}. For a more in-depth overview of load-balancing strategies in MD see sec. 5.3 of \citep{thompson2022lammps}.

An alternative approach to the MPI programming model are asynchronous many-task runtime systems \cite{RaicuFZ08}.
The number of tasks in many-task computing is typically significantly bigger than the number of cores. The many-task scheduler can therefore simply resolve load-imbalances by moving some tasks from busy to idle cores or even by moving tasks across nodes. 
One example is the Charm++ asynchronous programming model \citep{kale2020charm++short} which serves as the underlying runtime system of the MD simulation package NAMD \citep{phillips2020scalable} and of a highly scalable implementation of the finite element method \citep{bakosi2021asynchronous}.
Another emerging candidate library is HPX, which is a C++ standards-compliant library that provides wait-free asynchronous execution of tasks and synchronization through futures \citep{kaiser2020hpx}.

We have integrated HPX into ESPResSo++ for load-balancing on a single node, whereas we currently kept MPI to communicate between nodes. The HPX evaluation starts by again investigating balanced a Lennart-Jones fluid and a polymer melt to understand the overheads of HPX. We then investigate an unbalanced spherical configuration in which all particles are within a circle. The evaluation shows that the HPX overhead can be bounded by 5\% for perfectly balanced simulations and that HPX can improve performance by a factor of 1.4 for our unbalanced setting. 

In this paper, we therefore present our efforts at modernizing ESPResSo++ by optimizing its data layout, enabling SIMD vectorization and integrating the capabilities for fine-grained parallelism offered by HPX.
Our aim is to evaluate the benefits of using HPX to provide node-level parallelism compared to traditional MPI-based parallelism.

This paper is structured as follows:
In Section \ref{sec:background}, we introduce some of the general concepts and calculations involved in molecular dynamics simulations and the particular design principles of ESPResSo++.
We also describe the details of the HPX library that are relevant to our work.
In Section \ref{sec:optimizations}, we illustrate the serial optimizations we performed on the data layout and critical loops of ESPResSo++ and show how we extended the pure MPI-based parallelization in ESPResSo++ to an MPI+HPX parallelization model.
Then, we evaluated the performance of these optimizations and show the results in Section \ref{sec:evaluation}.
We list related publications in Section \ref{sec:related-work}.
Finally, we conclude our work in Section \ref{sec:summary-discussion} and discuss possible extensions to distributed parallelism in Section \ref{sec:future-work}.

\section{Background}
\label{sec:background}

In this section, we will recall the standard methods of performing molecular dynamics simulations on parallel machines.
We will also discuss particular features and design principles of the target MD application, ESPResSo++.
Finally, we will introduce the asynchronous many-task programming paradigms that will be used in the parallel optimizations of ESPResSo++.

\subsection{Molecular dynamics simulations}

\begin{figure}[h!]
    \centering
    \includegraphics[width=0.6\columnwidth]{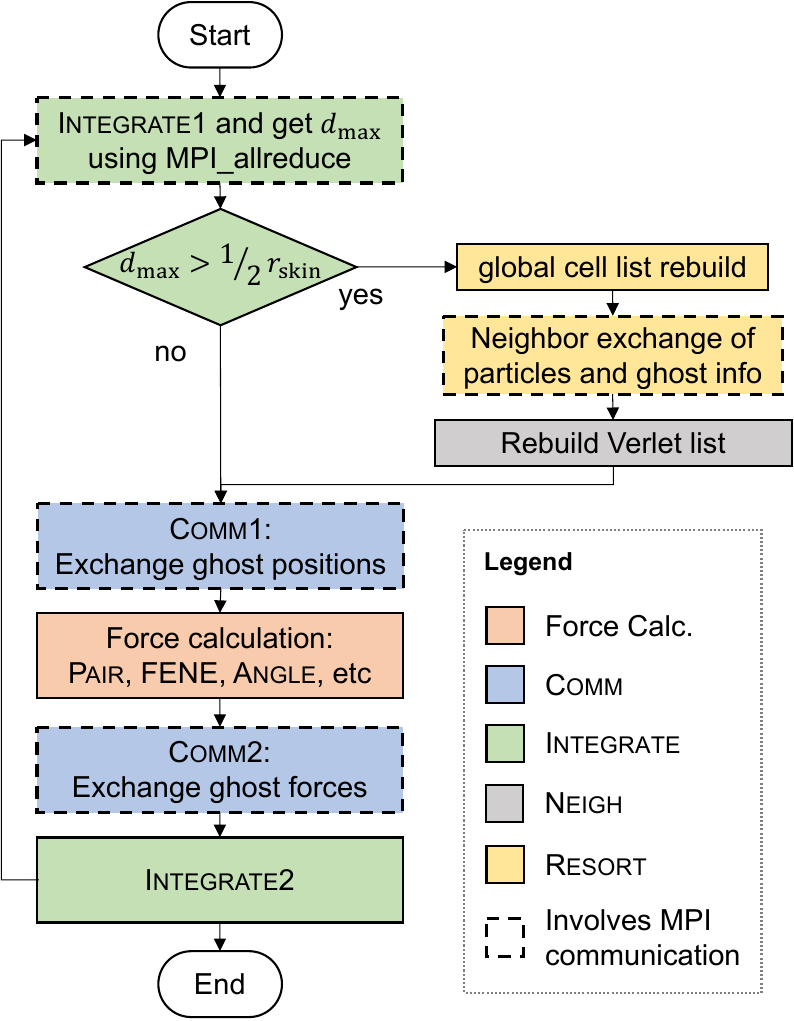}
    \caption{Velocity Verlet integration scheme used by ESPResSo++ which updates particles' velocities and positions in two half-steps (\textsc{Integrate1} and \textsc{Integrate2}) and uses a skin size $r_\text{skin}$ to determine the frequency of particle re-binning done by \textsc{Resort}. Colors indicate groupings of different sections together while dashed line borders (\texttt{---}) indicate sections that involve MPI communication.}
    \label{fig:integration-flowchart}
\end{figure}

Molecular dynamics (MD) simulations are used to model the dynamical properties of interacting particles in a system \citep{allen2017computer}.
The central task of MD simulations is to integrate Newton's equations of motion
\begin{equation}
    m_i\cdot\ddot{\mathbf{r}}_i=\mathbf{F}_i
    \label{eq:newton}
\end{equation}
where $m_i$ is the mass of the particle $i$, $\mathbf{r}_i$ is its position vector and $\mathbf{F}_i$ is the total force on $i$ from interactions with all other particles in the simulation.
These forces can be derived from a potential energy $U$ that depends on the coordinates of all $N$ particles $\mathbf{r}^N$ such that $\mathbf{F}=-\nabla U$.
The forces may include pairwise interactions, three-body interactions and higher order many-body interactions.

As shown in Figure \ref{fig:integration-flowchart}, the simulation evolves in an iterative process in which the positions are used to calculate the forces, which are in turn used to update the velocities and positions for the next time step.
For the latter, known as the \textsc{Integrate} step, numerical integrators such as the Velocity Verlet method are commonly employed \citep{verlet1967computer,swope1982computer}.

\subsubsection{Short-range non-bonded potentials}

Short-range force models comprise the most common class of interactions that are used in MD simulations.
They physically result from electronic screening effects of long-range interactions and they also reduce computational load by restricting force contributions to a smaller region around each particle \citep{plimpton1995fast}.

One commonly used example is the Lennard-Jones (LJ) potential which is given by
\begin{align}
    V(r) = 4\varepsilon\left[
        \left( \frac{\sigma}{r} \right)^{12} - \left( \frac{\sigma}{r} \right)^{6}
        \right]
\label{eq:lj}
\end{align}
where $\sigma$ and $\varepsilon$ are parameters that describe the size of the particle and the depth of the potential well, respectively.
If, like in the example of Equation \eqref{eq:lj}, the interaction decreases sufficiently rapidly to zero, the search for interacting particle pairs can be confined to within a distance $r<r_\text{cut}$ in order to calculate the total force on every particle.

\subsubsection{Neighbor lists}
Determining which particles are within cutoff of each other can be facilitated by binning the particles into cubic cells of minimum length $r_\text{cut}$.
Then the search for possible interaction pairs could be restricted to its cell and the 26 surrounding neighbor cells.
For interactions that obey Newton's third law, the symmetry of force calculations, $\vec{F}_{ij} = -\vec{F}_{ji}$, can be used to reduce the search to 13 neighbor cells.
The interacting particles may then be recorded in a list of pairs known as a Verlet list \citep{verlet1967computer}.
To reduce the computational cost of building the Verlet list, an additional buffer of thickness $r_\text{skin}$ is added to the cell size such that $r_\text{cell} \geq r_\text{cut} + r_\text{skin}$.
This allows some particles to move out of the neighborhood for some time steps without needing to reassign particles into cells and to recompute the Verlet lists frequently.

\subsubsection{Spatial decomposition}

The most common way to parallelize molecular dynamics simulations on distributed-memory machines is by subdividing the simulation box into smaller \textit{nodes}, each assigned to one processor \citep{plimpton1995fast}.
Every processor takes care of computing the forces and updating positions and velocities of particles in its own node.
Particles are then allowed to enter and exit a node by reassigning them to a different cell and node during the so-called \textsc{Resort} step.

Each node is composed of \textit{real cells} that belong to the spatial domain of that node and a surrounding layer of \textit{ghost cells} that contain copies of particles from neighboring nodes \citep{halverson2013espressopp}.
The ghost cells are needed to correctly account for all force contributions at the node boundaries and they are updated at each time step using local message-passing communication, referred to as the \textsc{Comm} step.

\subsection{ESPResSo++}

Started in 2008 as a joint collaboration between the Fraunhofer SCAI Institute and the Theory Group of the Max Planck Institute for Polymer Research, ESPResSo++ has become a central tool to perform coarse-grained simulations and to provide a sophisticated and versatile framework for the development of new methods and algorithms \citep{halverson2013espressopp, guzman2019espresso++}.
In order to simplify data exchange in complex workflows in combination with other software packages such as LAMMPS \citep{thompson2022lammps}, GROMACS \citep{abraham2015gromacs} and VOTCA \citep{ruhle2009versatile, wehner2018electronic}, interfaces to these programs have been implemented.
ESPResSo++ supports a variety of standard MD algorithms (e.g. Velocity Verlet Integrator in NVE, NVT, NPT Ensembles) as well as several advanced methods (e.g. AdResS, H-AdResS, equilibration of polymer melts, Lattice Boltzmann, 3-body non-bonded Stillinger-Weber or the Tersoff potential) and modern data structures (e.g., H5MD) in
parallel file I/O environments.
More recent and ongoing developments include the integration of the ScaFaCoS library\footnote{\url{http://www.scafacos.de}} for long range interactions and Lees-Edwards boundary conditions \citep{xu2021implementation}.
Collaborative development within an international group of scientists, including continuous integration (CI) and unit testing, happens on the GitHub platform.\footnote{\url{https://www.github.com/espressopp}}

\subsection{Asynchronous many-task programming models}

Most applications that run on modern parallel architectures rely on the MPI+X programming model \citep{wolfe2014compilers} in which MPI handles inter-node communication and ``X'' is a hardware-dependent node-level programming paradigm such as OpenMP\footnote{\url{https://www.openmp.org/wp-content/uploads/HybridPP_Slides.pdf}}, CUDA\footnote{\url{https://developer.nvidia.com/blog/introduction-cuda-aware-mpi/}}, and OpenACC\footnote{\url{{https://www.openacc.org/sites/default/files/inline-images/Specification/OpenACC-3.1-final.pdf}}}, or a performance-portability framework such as Kokkos \citep{trott2022Kokkos} and RAJA \citep{beckingsale2019raja}.
Solutions like QUO have also been implemented to handle thread-level heterogeneity while ensuring full utilization of CPU cores \citep{gutierrez2017accommodating}.

An emerging class of new parallel programming paradigms are asynchronous many-task (AMT) runtime systems which can provide fine-grained parallelism that can handle a large number of threads and efficiently distribute work across a node.
One such example is HPX.
The High Performance ParalleX (HPX) library is a C++ runtime system that can handle fine-grained parallelism in modern architectures \citep{kaiser2020hpx}.
It allows users to write code based on a task dependency graph and takes care of the thread scheduling and execution of tasks and communication between distributed compute nodes.

We specifically focus on the thread management aspects of HPX to write and execute multithreaded code.
This is achieved through the concept of \textit{futurization}.
HPX provides an asynchronous return type \texttt{hpx::future<T>} which hides the execution of function calls and returns immediately even though the function has not completed its execution.
The futures can also be consumed by local control objects (LCOs) such as \texttt{hpx::future<T>::then}, \texttt{hpx::wait\_all} and \texttt{hpx::shared\_future}, which enable the flow of task dependencies.
The API also provides high-level parallel algorithms aligned to current and proposed C++ standards and extends them with asynchronous versions \citep{kaiser2020hpx}.

In the distributed case, HPX also provides an Active Global Address Space (AGAS) which registers objects with global identifiers that allows remote function calls while hiding explicit message passing.
However, since we are focusing on node-level optimizations, AGAS will not be used in this work and inter-node communication will be done through MPI.

\section{Optimizations}
\label{sec:optimizations}

In this section, we discuss the key steps taken to improve the performance of ESPResSo++.
We start with baseline optimizations that involve transforming the data layout and ensuring that use of SIMD vectorization is maximized by the compiler.
Then, we implement thread-level parallelism by integrating the HPX runtime system into ESPResSo++.

The standard implementation of ESPResSo++ is heavily dependent on the \texttt{Particle} data structure.
Since we want to maintain compatibility with the analysis tools in the current implementation while taking advantage of the performance improvements, we have chosen to implement the following optimizations as additional submodules within the \texttt{espressopp} Python module.
The \texttt{vec} submodule contains the data structures for improved particle data layout described in Section \ref{subsec:improved-data-layout} and the vectorized routines for force calculation and integration described in Section \ref{subsec:vectorization}.
On the other hand, the \texttt{hpx4espp} submodule contains the corresponding routines that use the node-level parallelism provided by HPX described in Section \ref{subsec:node-level}.
This means that once the corresponding optimized versions of subroutines have been implemented, users only need to modify their python scripts by indicating the submodule.
For example, the original version of the FENE potential is instantiated by
\texttt{espressopp.interaction.FENE},
while for the vectorized version the \texttt{vec} submodule is specified resulting in\\
\texttt{espressopp.vec.interaction.FENE}.
The introduction of these submodules is facilitated by the extensibility and object-oriented design of ESPResSo++.

\subsection{Improved data layout}
\label{subsec:improved-data-layout}

\begin{figure}
    \centering
    \begin{subfigure}{\columnwidth}
        \centering
        \includegraphics[width=0.75\columnwidth]{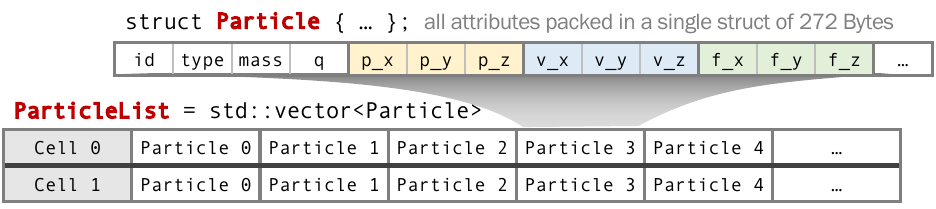}
        \caption{Original layout in which each \texttt{Particle} struct contains 272 bytes and every cell has a \texttt{vector} of these structs.}
        \label{subfig:data-layout-orig}
    \end{subfigure}
    \vspace{1em}

    \begin{subfigure}{\columnwidth}
        \centering
        \includegraphics[width=0.3\columnwidth]{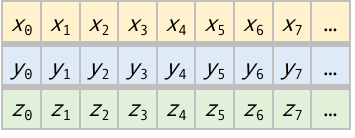}
        \caption{Improved layout using structure of arrays (SoA) in which each attribute is stored as a separate array.}
        \label{subfig:data-layout-soa}
    \end{subfigure}
    \caption{Original and improved layout for particle data.}
\end{figure}

Figure \ref{subfig:data-layout-orig} shows the particle data layout in the standard implementation of {ESPRes\-So++}. All data for a single particle are stored in a large struct of 272 bytes called \texttt{Particle} including basic attributes such as type, position, velocity, force, and other attributes required for more complex simulations.
The particles are further grouped into cells, each containing its own \texttt{std::vector<Particle>}.
This layout was chosen since it provides the most straightforward way of accessing particle data.
This makes it compatible with the goal of extensibility and the object-oriented design of ESPResSo++, and it is well suited for algorithms working on nearly all particle data \citep{halverson2013espressopp}.
However, most of the time-consuming operations, such as force calculation and neighbor search, rely on only a few attributes.
Performing those operations with this layout requires strided memory accesses through different attributes and operations with this layout usually cannot be auto-vectorized by compilers.

In order to optimize these operations, we employ an alternative layout in the form of a structure of arrays (SoA) in which every attribute of type \texttt{T} is stored in its own separate \texttt{std::vector<T>} as shown in Figure \ref{subfig:data-layout-soa}.
SoA is known to improve cache re-use and is amenable to compiler-assisted vectorization of time-critical loops \citep{watanabe2019simd}.

To maintain cache coherence and prevent false sharing during multithreaded accesses, the member arrays are allocated aligned to 64-byte boundaries.
For \texttt{std::vector}s this was done by using the \texttt{aligned\_allocator} provided by the Boost.Align library\footnote{\url{https://www.boost.org/}}.
The entries between the end of one cell and the start of the next cell are also padded with dummy particles that lie far away from the simulation box to ensure that the entries for the next cell are also properly aligned.

The SoA layout is utilized only during the \textsc{Integrate} and force calculation steps while the original layout is used in the \textsc{Resort} stage since it requires all particle attributes.
Thus, to ensure accuracy, the data between them are synchronized at the beginning and end of the simulation, and before and after the intermediate \textsc{Resort} stages.

\subsection{Vectorization}
\label{subsec:vectorization}

In this paper, we are specifically interested in optimizing for the vectorization capabilities provided by AVX-512 on Intel Skylake and Ice Lake processors.
However, we also want to keep critical loops portable to other compilers and CPU architectures while taking advantage of the SoA data layout.
Thus, we mainly rely on compiler auto-vectorization to ensure that SIMD instructions are efficiently utilized.
Further directives were added in the code to inform the compiler of alignment and to assert the absence of data dependencies among vector entries \citep{jeffers2016intel}.

\begin{figure}
    \centering
    \begin{subfigure}[t]{\columnwidth}
        \centering
        \includegraphics[width=0.6\columnwidth]{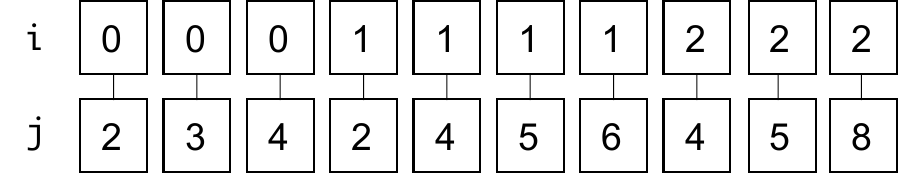}
        \caption{List of pairs representing the indices of or pointers to interacting particles used in the standard implementation of ESPResSo++.}
        \label{subfig:vl-list-of-pairs}
    \end{subfigure}
    \vspace{2em}

    \begin{subfigure}[t]{\columnwidth}
        \centering
        \includegraphics[width=0.6\columnwidth]{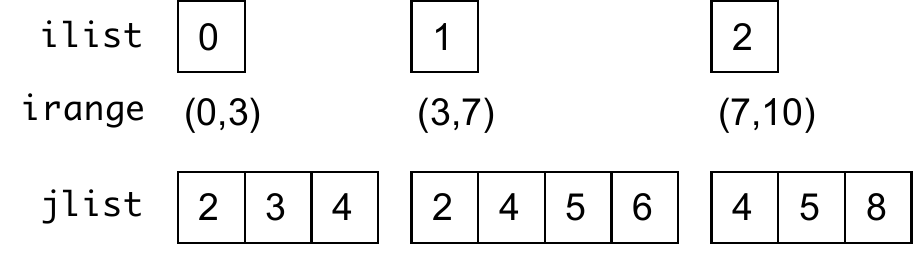}
        \caption{\textsc{SortedList} based on \cite{watanabe2019simd} which groups together the pairs that have the same first particle.}
        \label{subfig:vl-sorted-list}
    \end{subfigure}
    \vspace{1em}

    \caption{Possible representations of a Verlet list.}
    \label{fig:vl}
\end{figure}

The most time-consuming sections of typical short-range non-bonded MD simulations are the neighbor list rebuild and the force calculation.
Particularly for pair potentials, these steps require an efficient representation of the Verlet list.
In the current version of ESPResSo++, it is represented as a list of pairs of pointers $(i,j)$ to the corresponding \texttt{Particle} structs as shown in Figure \ref{subfig:vl-list-of-pairs}.

In our work, we use a more compact and vector-friendly representation known as a \textsc{SortedList} \citep{watanabe2019simd}. Here the  $j$-particles that interact with the same $i$-particle are grouped together and stored in a contiguous list as shown in Figure \ref{subfig:vl-sorted-list}.
Thus, during force calculations the \textsc{SortedList} is traversed using a double for-loop: first, through the array of indices \texttt{ilist} and ranges \texttt{irange}, and then through the entries of \texttt{jlist} that are contained within the range.
Since every $i$ comes with distinct $j$ entries, this structure leads to the possibility of applying vectorization on the inner for-loop over $j$.

\subsection{Node-level parallelism using HPX}
\label{subsec:node-level}

\begin{figure}[t!]
    \centering
    \includegraphics[width=0.75\columnwidth]{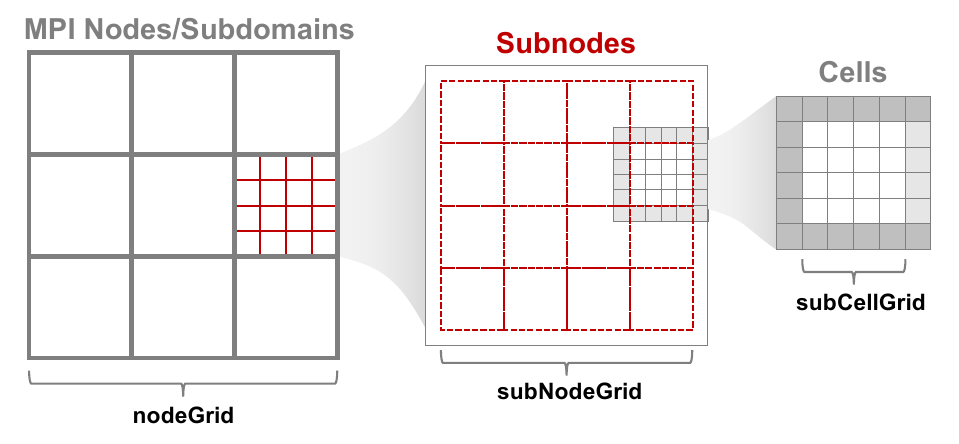}
    \caption{MPI nodes are further divided into subnodes which determine the task granularity. }
    \label{fig:subnodes}
\end{figure}

To enable multithreaded parallelism with HPX, we further divide the nodes into smaller \textit{subnodes} as shown in Figure \ref{fig:subnodes}. All computation needed for one subnode is packaged into one HPX task.
Each subnode contains only the particle data of the cells that belong to it.
This allows some operations such as the integration of positions and velocities to be done in parallel for each subnode without needing explicit locking of resources.
However, force calculation requires modifying the force values of two or more particles that may reside in different subnodes.
To avoid race conditions in the multithreaded version, Newton's third law is not used for interactions between particles belonging to different subnodes so that their force values are calculated separately.
This allows the force calculation for each subnode to be run concurrently but at the expense of some redundant force calculations at the subnode boundaries.

The number of cells within the subnode grid determines the task granularity.
If the subnode is too small, execution would be dominated by overheads, in particular ghost layer exchange.
In contrast, a large grid size results in fewer subnodes which hinders HPX's work stealing capabilitiest.
Thus, the number of subnodes per core, known as an oversubscription factor, has to be tuned in order to find the optimal point between overheads and starvation \citep{bremer2019performance,grubel2015performance}.
This tuning procedure could be done manually by performing several runs of a few time-steps while varying the number of subnodes at each run, starting with the number of threads per MPI locality until no further decrease in elapsed time is recorded.
This optimal number of subnodes will vary for every simulation system depending on the task size and relative load imbalances among the resulting subnodes with the assumption that the relative load distribution does not vary too much during the simulation.
With this approach we can leverage the HPX work stealing capabilities to achieve load balancing between the threads of one MPI rank. Since the number of tasks is in general higher than the number of threads new tasks get assigned to a thread as soon as it finishes its work.

To execute the MD operations in parallel across multiple threads, we rely on the C++ standards-compliant parallel algorithms provided by HPX.
For the parallel execution of plain loops over subnodes we use \texttt{hpx::parallel::for\_loop} and for those involving reductions we use \texttt{hpx::parallel::transform\_reduce}.
Their signatures are similar to the C++ Standard Template Library (STL) algorithms of the same name except that the first argument in HPX requires an execution policy which indicates whether and how to perform the algorithm in parallel \citep{kaiser2020hpx}.
Execution policies are gradually being adopted into the C++ standard, so that calls to functions in the \texttt{hpx} namespace may be replaced with \texttt{std} in the future.

\section{Performance evaluation}
\label{sec:evaluation}

Evaluation was performed on three different computing facilities: the Mogon II supercomputer at the Johannes Gutenberg University Mainz which is equipped with Intel Skylake processors, the Raven HPC System at the Max Planck Computing and Data Facility (MPCDF) which has the newer Intel Ice Lake processors, and an AMD EPYC 7452 cluster also at the MPCDF.
Hardware specifications and compilers that were used are described in Table \ref{tab:specs}.

We used the following software versions: HWLOC 2.2, Boost 1.72.0, Python 3.8 and HPX 1.5.1 \citep{kaiser2020hpx151short}.
Binaries were compiled with Intel C++ Compiler with the flags ``-O3 -xHost -qopt-zmm-usage=high -restrict" to enable full AVX-512 optimizations on Mogon and Raven, and with GCC compiler with flags ``-O3 -march=native'' on the AMD cluster.

\begin{table}[t!]
    \centering
    \begin{tabularx}{\linewidth}{l|
        >{\centering\arraybackslash}X
        >{\centering\arraybackslash}X
        >{\centering\arraybackslash}X}

    \toprule
    Name & \textbf{Mogon} & \textbf{Raven} & \textbf{AMD} \\
    \midrule
    Processor
        & Intel Gold 6130 (Skylake)
        & Intel Xeon IceLake-SP 8360Y
        & AMD EPYC 7452 \\
    Base frequency & 2.10 GHz & 2.40 GHz & 2.35 GHz \\
    Sockets per node & 2 & 2 & 2 \\
    Cores per socket & 16 & 36 & 32 \\
    Vector Extensions & AVX-512 & AVX-512 & AVX2 \\
    Compiler & ICC 19.1.1.217 & ICC 2021.3.0 & GCC 11.2.0 \\
    \bottomrule
    \end{tabularx}
    \caption{Specifications of the compute nodes used for performance evaluation.}
    \label{tab:specs}
\end{table}

Two simulation systems were used to evaluate performance -- Lennard-Jones and polymer melts.
The quantities used here will be expressed in Lennard-Jones dimensionless units with $m=\varepsilon = \sigma = 1$ which also results in reduced units of time since $(\varepsilon/m\sigma^2)^{1/2}=1$ \citep{allen2017computer}.

The Lennard-Jones simulation was initialized with particles in a cubic lattice at a density $\rho=0.8442$, and a cutoff distance of $r_\text{cut}=2.5$ and a buffer thickness of $r_\text{skin}=0.3$ were set.
A Langevin thermostat was introduced to equilibrate the particles to some target temperature $T$.

We also prepared a polymer melt simulation containing ring polymers of chain length 200 and density $\rho=0.85$.
A repulsive Lennard-Jones interaction exists between all monomers within a cutoff distance $r_\text{cut}=2^{1/6}$ and skin size $r_\text{skin}=0.4$.
Aside from the short-range non-bonded interactions, the polymer melt simulation includes bonded interactions composed of a FENE potential between pairs along the chains and a cosine potential on triples that form angles \citep{kremer1990dynamics}.

The simulations were executed with a fixed step size $\Delta t = 0.005$.
We measured the elapsed time of the integrator loop and excluded any initialization steps such as setting up classes and reading particle data.
Within this period, the timings of the following key sections were also collected:
\begin{itemize}
    \item Forces - calculating non-bonded (LJ/\textsc{Pair}) and bonded (FENE, Angle) interactions
    \item \textsc{Comm} - communicating positions and collecting forces for ghost layers
    \item \textsc{Integrate} - updating positions and velocities
    \item \textsc{Neigh} - rebuilding neighbor lists
    \item \textsc{Resort} - sorting particles to cells and nodes, and copying particle data between original and SoA layout
\end{itemize}
These correspond to the groupings of sections described in Figure \ref{fig:integration-flowchart} in which timers for sections with the same color are combined (e.g. $\textsc{Comm1} + \textsc{Comm2} = \textsc{Comm}$).
The values presented are averages across all MPI ranks.

\subsection{Baseline optimizations}

\begin{figure}
    \centering
    \begin{subfigure}[t]{0.42\textwidth}
        \centering
        \includegraphics[width=\textwidth]{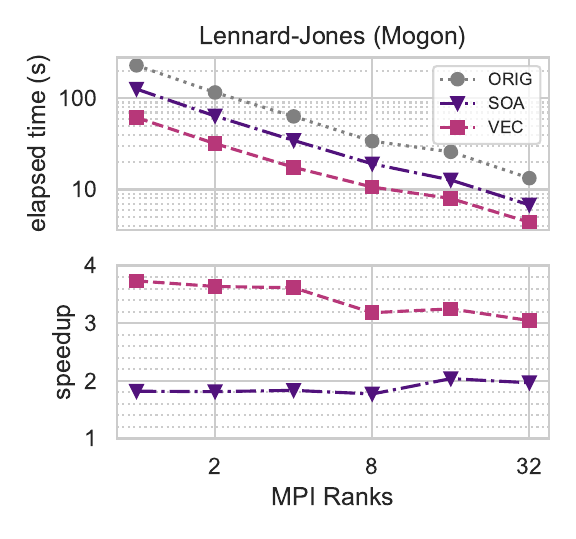}
        \caption{LJ fluid on Mogon}
        \label{subfig:soa-vec-run-lj-mogon}
    \end{subfigure}
    \begin{subfigure}[t]{0.42\textwidth}
        \centering
        \includegraphics[width=\textwidth]{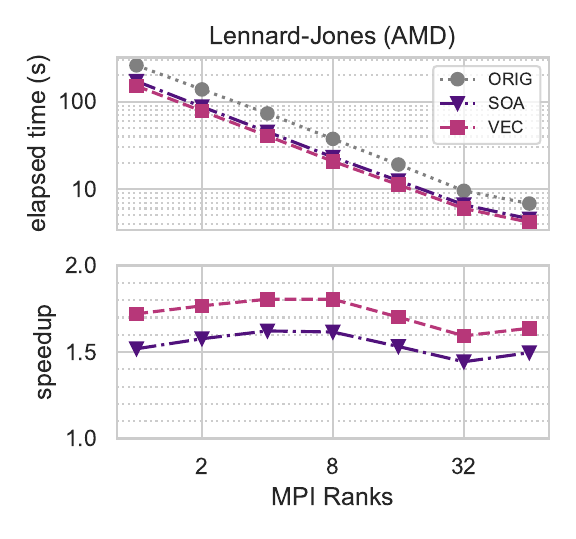}
        \caption{LJ fluid on AMD node}
        \label{subfig:soa-vec-run-lj-amd}
    \end{subfigure}

    \begin{subfigure}[t]{0.42\textwidth}
        \centering
        \includegraphics[width=\textwidth]{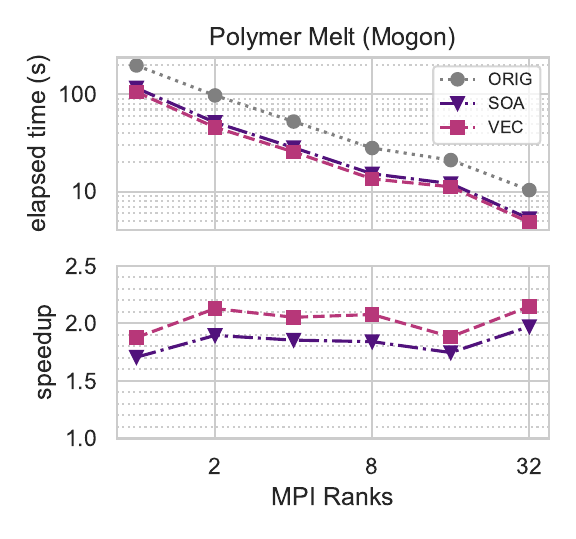}
        \caption{Polymer melt on Mogon}
        \label{subfig:soa-vec-run-pm-mogon}
    \end{subfigure}
    \begin{subfigure}[t]{0.42\textwidth}
        \centering
        \includegraphics[width=\textwidth]{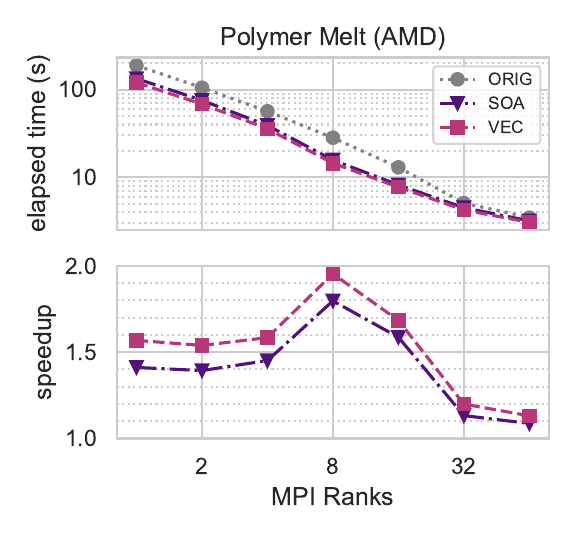}
        \caption{Polymer melt on AMD node}
        \label{subfig:soa-vec-run-pm-amd}
    \end{subfigure}
    \begin{subfigure}[t]{0.42\textwidth}
        \includegraphics[width=\textwidth]{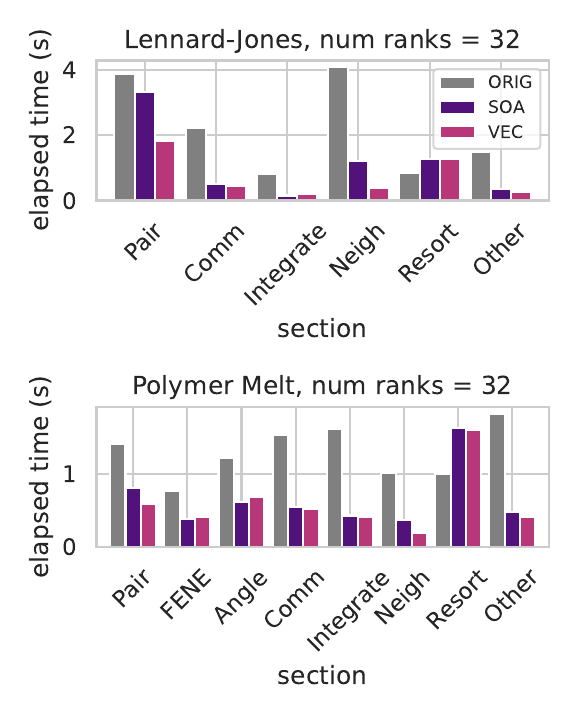}
        \caption{Sectional time on Mogon.}
        \label{subfig:soa-vec-sec-mogon}
    \end{subfigure}
    \begin{subfigure}[t]{0.42\textwidth}
        \includegraphics[width=\textwidth]{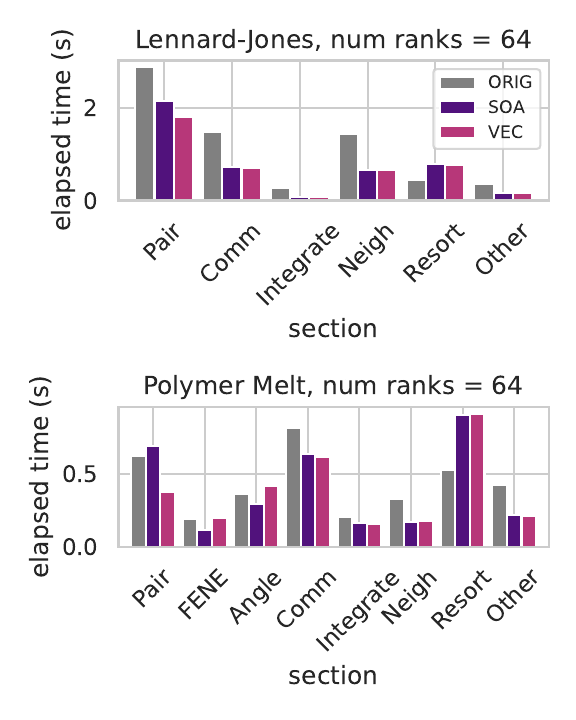}
        \caption{Sectional time on AMD Node}
        \label{subfig:soa-vec-sec-amd}
    \end{subfigure}

    \caption{Strong scaling behavior on one node of Mogon and AMD. Speedups shown are calculated with respect to \textsc{orig} at the same number of MPI ranks. A higher speedup is obtained for the LJ fluid due to its pair potential having a larger cutoff distance than the one used for the polymer melt simulation. For elapsed time, lower is better.}
    \label{fig:soa-vec-run-mogon}
\end{figure}

We first evaluate the performance benefits following the transformation of the particle data layout to SoA and the vectorization of critical loops.
To do this, we compare three cases:
\begin{itemize}
    \item \textsc{orig} - the original implementation,
    \item \textsc{soa} - the implementation optimized with SoA particle data layout but with auto-vectorization disabled using the flags ``-no-vec -no-simd'' on ICC and ``{-fno-tree-vectorize}'' on GCC, and
    \item \textsc{vec} - the fully vectorized implementation.
\end{itemize}
We performed the measurements on a single node on each of the three machines listed in Table \ref{tab:specs} using the Lennard-Jones fluid and polymer melt simulation as benchmark cases.
We present the timing results in Figure \ref{fig:soa-vec-run-mogon} and Figure \ref{fig-si:soa-vec-run-raven}.

As our first benchmark case, we simulate a Lennard-Jones fluid containing $N=262,144$ particles running for 1000 time steps and equilibrated to $T=1.0$.
The overall elapsed time and the speedup with respect to \textsc{orig} are shown in Figures \ref{subfig:soa-vec-run-lj-mogon}-\ref{subfig:soa-vec-run-lj-amd}.
On Mogon, we observe a $2\times$ speedup from \textsc{orig} to \textsc{soa} due to the change in data layout, and a further $1.5\times$ speedup from \textsc{soa} to \textsc{vec} due to the vectorization of the Lennard-Jones interaction and the neighbor list rebuild.
This is reduced on the AMD node to $1.5\times$ and $1.09\times$ respectively since the processor has AVX2 extensions which has a smaller vector width and reduced vectorization capabilities.

The elapsed time of different sections on a full node are shown in Figures \ref{subfig:soa-vec-sec-mogon}-\ref{subfig:soa-vec-sec-amd}.
Speedups can be observed in all sections that use the SoA layout.
There is an overhead in the \textsc{Resort} section due to the additional copying of data from the SoA layout to the unoptimized layout on which the re-binning is performed, but this is compensated by the larger speedups in the other sections.
Among the different code sections, speedups resulting from vectorization are more significant in the force calculation and neighbor list rebuilds because these sections involve a traversal of the \textsc{SortedList} which has a higher trip count than the iteration through the particles in the other sections.

We also performed the same comparison on a polymer melt simulation with $N=320,000$ particles and show the results in Figures \ref{subfig:soa-vec-run-pm-mogon}-\ref{subfig:soa-vec-run-pm-amd}.
On Mogon, a $2\times$ speedup is achieved from \textsc{orig} to \textsc{soa}, similar to the LJ simulation.
However, the speedup from \textsc{soa} to \textsc{vec} was reduced to only $1.07\times$.
This is because the smaller cutoff for the pair potential results in fewer neighbors per particle at an average of 9.4 compared to 41.2 for the LJ setup, which in turn results in fewer iterations for the vectorized inner loop of the \textsc{SortedList} traversal in \textsc{Pair} and \textsc{Neigh}.
Also, the other forces --- FENE and Angle --- were not vectorized in the same manner since they require mechanisms for conflict detection which are not yet supported by auto-vectorization.
For the simulation on Raven, speedup benchmarks for both systems are similar to those on Mogon and are hence not further discussed in this subsection (See the supplementary information for more details).

The speedups we obtained from the \textsc{soa} implementation to the \textsc{vec} implementation can be compared with the ideal speedup which can be calculated by assuming perfect vectorization of the \textsc{Pair} and \textsc{Neigh} operations.
If we denote the execution time for the non-vectorized sections as $t_\text{rest}$, we can quantify the maximum speedup as
\begin{equation}
    S_\text{max} =
        \dfrac{t_\text{rest} + t_\textsc{Pair} + t_\textsc{Neigh}}
        {t_\text{rest} + \left({t_\textsc{Pair} + t_\textsc{Neigh}}\right)/{W}}
\label{eq:vec-perf-model}
\end{equation}
where the execution times $t_\textsc{Pair}$ and $t_\textsc{Neigh}$ are those obtained with \textsc{soa}, and $W$ denotes the vector width or how many double-precision floating point numbers can fit in the register of the vectorized instructions (8 for AVX-512 and 4 for AVX2).

We apply Equation \eqref{eq:vec-perf-model} to the data in Figures \ref{subfig:soa-vec-sec-mogon}-\ref{subfig:soa-vec-sec-amd}, and present the results in Table \ref{tab:vec-perf-model}.
On Mogon, we find that for the Lennard-Jones simulation $S^\text{LJ}_\text{max} = 2.39$ compared to our measured $S^\text{LJ}=1.55$ and for the polymer melt simulation $S^\text{PM}_\text{max} = 1.24$ whereas we obtained $S^\text{PM} = 1.09$ in our implementation.
This difference is mainly due to the fact that the \textit{j}-particles in the \textsc{SortedList} are not contiguous and have to be gathered before being simultaneously processed by the SIMD instructions.
Similar values were obtained on Raven (Figure \ref{fig-si:soa-vec-run-raven}c) since its Ice Lake processor has the same AVX-512 vectorization features as Skylake on Mogon.
On the other hand, the speedups obtained on the AMD machine were relatively reduced due to the shorter SIMD width and the fewer vectorization capabilities of the AVX2 instruction set.

\begin{table}[h!]
\caption{Actual speedups ($S$) from \textsc{soa} to \textsc{vec} and corresponding ideal speedups ($S_\text{max}$) calculated from \eqref{eq:vec-perf-model}}
\centering
\begin{tabular}{l|c|cc|cc}
\toprule
 & $W$ & $S^\text{LJ}$ & $S_\text{max}^\text{LJ}$ & $S^\text{PM}$ & $S_\text{max}^\text{PM}$ \\
\midrule
Mogon & 8 & 1.55 & 2.39 & 1.09 & 1.24 \\
Raven & 8 & 1.65 & 2.07 & 1.09 & 1.25 \\
AMD   & 4 & 1.09 & 1.86 & 1.04 & 1.26 \\
\bottomrule
\end{tabular}
\label{tab:vec-perf-model}
\end{table}

\begin{figure}[h!]
    \centering
    \includegraphics[width=0.75\columnwidth]{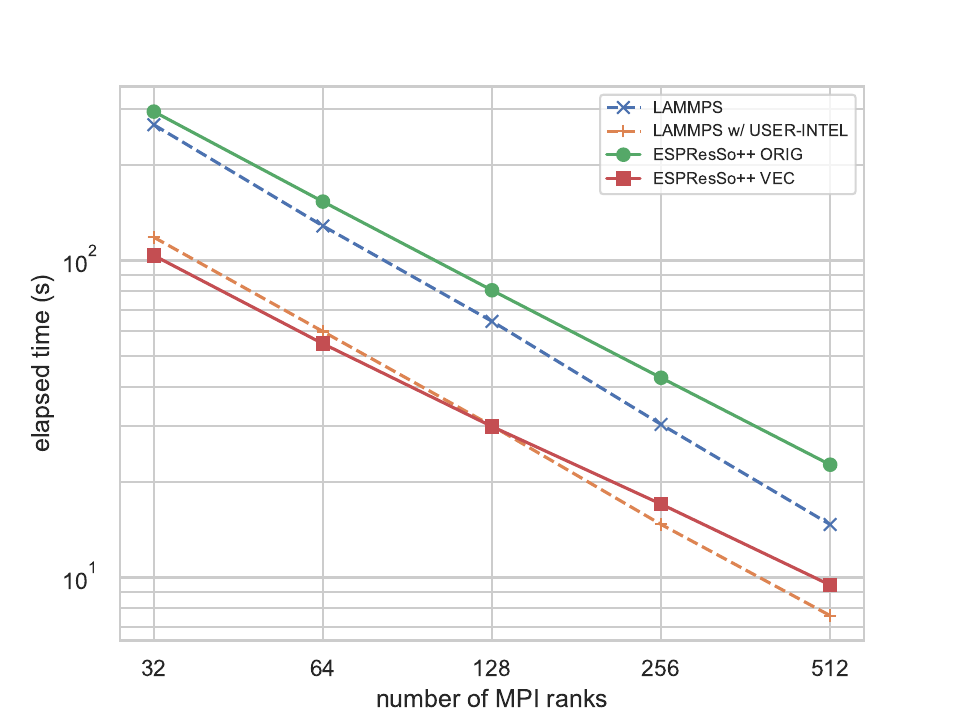}
    \caption{Strong scaling comparison with LAMMPS on Mogon.}
    \label{fig:vec-strong-scaling}
\end{figure}

We also compare our vectorized implementation with LAMMPS which provides an Intel-optimized USER-INTEL package \citep{lammpsstable29Oct2020}.
We ran a strong scaling comparison against the original implementation and fully vectorized version of ESPResSo++ and present the results in Figure \ref{fig:vec-strong-scaling}.
We find a similar performance of LAMMPS USER-INTEL and our optimized version of ESPResSo++ with the latter being faster up to 128 MPI ranks.
However, scalability is eventually limited by the \textsc{Resort} stage which is still based on the original implementation and could be further improved in future implementations.
Nevertheless, we have shown that even our SIMD optimizations already resulted in comparable overall performance with LAMMPS.

\subsection{MPI vs HPX comparison}
\label{subsec:mpi-vs-hpx}

\begin{figure}[h!]
    \centering
    \begin{subfigure}[t]{0.48\textwidth}
        \centering
        \includegraphics[width=\textwidth]{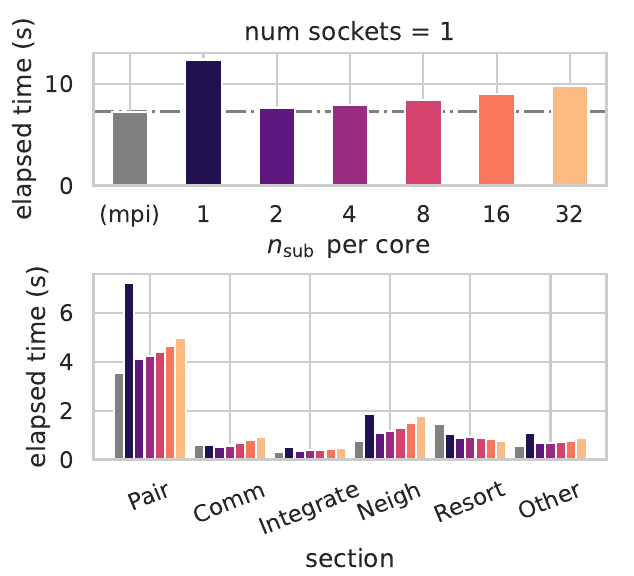}
        \caption{Mogon (1 MPI rank, 16 HPX threads)}\
        \label{subfig:mpi-vs-hpx-bulk-lj-nsock-1-mogon}
    \end{subfigure}
    \begin{subfigure}[t]{0.48\textwidth}
        \centering
        \includegraphics[width=\textwidth]{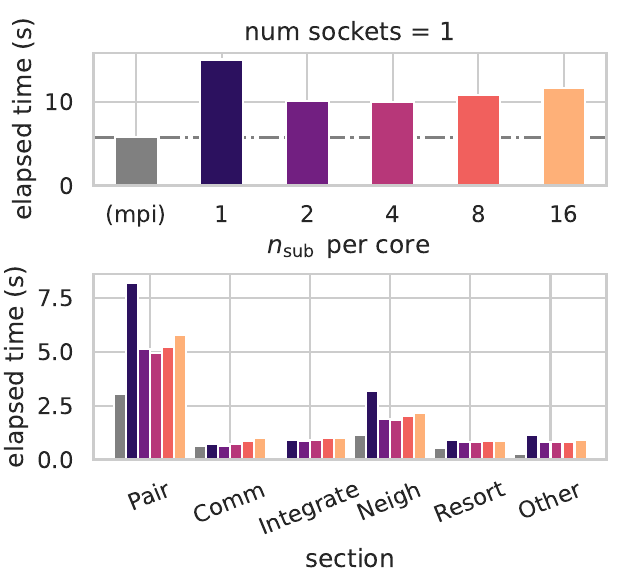}
        \caption{AMD (1 MPI rank, 32 HPX threads)}\
        \label{subfig:mpi-vs-hpx-bulk-lj-nsock-1-amd}
    \end{subfigure}

    \begin{subfigure}[t]{0.48\textwidth}
        \centering
        \includegraphics[width=\textwidth]{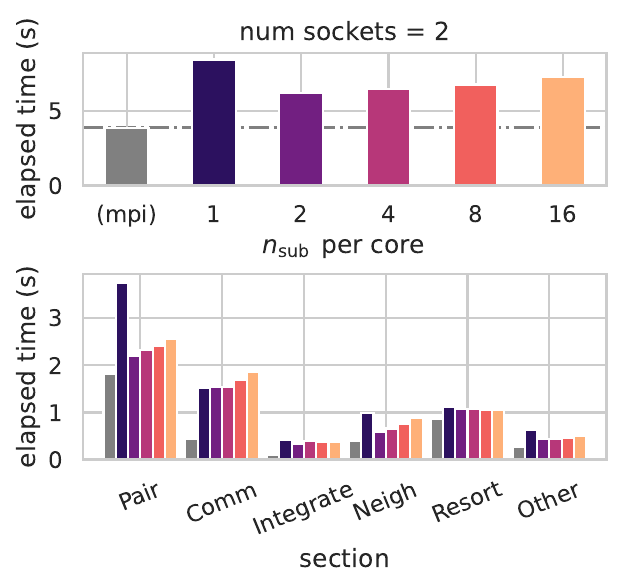}
        \caption{Mogon (2 MPI ranks, 16 HPX threads)}\
        \label{subfig:mpi-vs-hpx-bulk-lj-nsock-2-mogon}
    \end{subfigure}
    \begin{subfigure}[t]{0.48\textwidth}
        \centering
        \includegraphics[width=\textwidth]{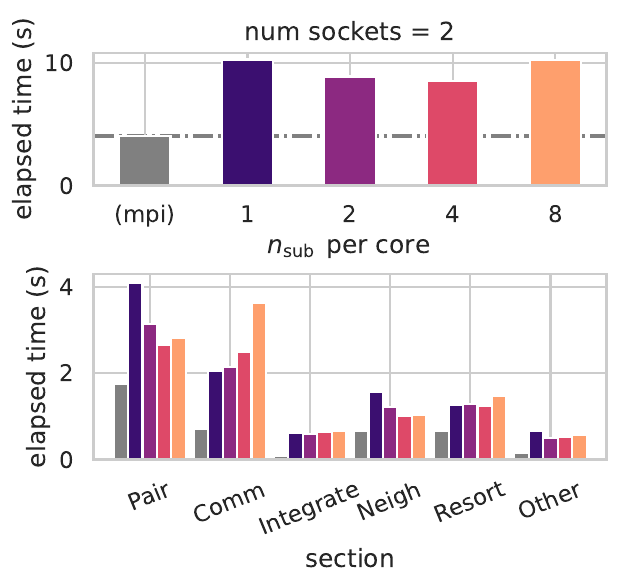}
        \caption{AMD (2 MPI ranks, 32 HPX threads)}\
        \label{subfig:mpi-vs-hpx-bulk-lj-nsock-2-amd}
    \end{subfigure}

    \caption{Elapsed time for the homogenous Lennard-Jones fluid with the MPI version ({\color{gray}-$\cdot$-$\cdot$}) and with HPX using different number of subnodes on Mogon.
    Lower time is better.
    On a single socket of Mogon, the performance of the HPX version at $n_\text{sub} = 32$ is similar to MPI. For the single node case, the performance is impacted by having only the main thread performing MPI calls. AMD also shows higher overhead for performing the \textsc{Pair} force calculation.}
    \label{fig:mpi-vs-hpx-bulk-lj}
\end{figure}

In this section, we evaluate our molecular dynamics implementation with HPX shared-memory parallelism in comparison to traditional MPI-based intra-node parallelism.
We specifically investigate two cases: a spatially homogenous simulation to measure the overheads resulting from the HPX implementation, and a spatially inhomogeneous simulation to demonstrate the load-balancing capabilities of HPX.
In particular, we focus on single node performance since we want to investigate the benefits from the work-stealing capabilities of HPX.
The MD simulations were executed on the compute nodes listed in Table \ref{tab:specs} equipped with two sockets.
For the HPX version, one MPI rank is assigned to each socket and the number of HPX threads for each rank is equal to the number of cores per socket.
For the baseline MPI version, one MPI rank with one thread is assigned to each physical core, which corresponds to the fully vectorized implementation evaluated in the previous section.

We start with a spatially homogenous case in which we expect the simulation to be dominated by implementation overheads.
We use the same Lennard-Jones setup from the previous section which contains $N=262,144$ particles and forms a cell grid of $(24,24,24)$.
For the HPX version, we perform the simulations for different numbers of subnodes per MPI rank, $n_\text{sub}$, starting from the number of cores on each socket and gradually increasing by a factor of two until no further subdivision is possible.
This autotuning procedure allows us to find the optimal value of $n_\text{sub}$.
For MPI version, we run only one $n_\text{sub}$ per MPI rank.

We first performed the evaluation on a single socket to isolate our measurements from the effects of MPI communication.
From the results shown in Figure \ref{subfig:mpi-vs-hpx-bulk-lj-nsock-1-mogon}, an immediate drop in elapsed time can be seen from one to two $n_\text{sub}$ per core followed by a steady slowdown as $n_\text{sub}$ continues to increase.
This means that executing two subnodes per thread is already sufficient since the \textsc{Pair} section does not improve anymore after this point.
On the other hand, as $n_\text{sub}$ increases, the number of force calculations that do not use Newton's third law also increases, which in turn increases the time for the \textsc{Pair} section.
At the optimal point for Mogon at two $n_\text{sub}$ per core, we find that the HPX implementation is $5\%$ slower compared to the traditional MPI implementation where the overhead originates from task sizes of the \textsc{Integrate} section being too small and the increase in necessary but redundant force calculations in the \textsc{Pair} section.
For AMD and Raven, we see from Figures \ref{subfig:mpi-vs-hpx-bulk-lj-nsock-1-amd}  and \ref{fig-si:mpi-vs-hpx-bulk-lj-raven}a that the optimal point is at four $n_\text{sub}$ per core and there is higher overhead from the \textsc{Pair} force calculation.

The same simulation was then performed on two MPI ranks assigned to two sockets with the same number of HPX threads for each rank.
From Figures \ref{subfig:mpi-vs-hpx-bulk-lj-nsock-2-mogon}-\ref{subfig:mpi-vs-hpx-bulk-lj-nsock-2-amd} we see similar behavior except for a uniform increase in communication time.
On Mogon, this results in a $37\%$ slowdown compared to running one MPI task per core on the entire compute node.
This is due to the fact that in the HPX implementation only the main thread in each MPI locality is performing communication so the number of communication calls is reduced and the data volume for each call is increased.
This can be overcome by future implementations of the distributed parallelism provided by HPX, which is beyond the scope of this paper.

\begin{figure}[t!]
    \centering
    \includegraphics[width=0.4\columnwidth]{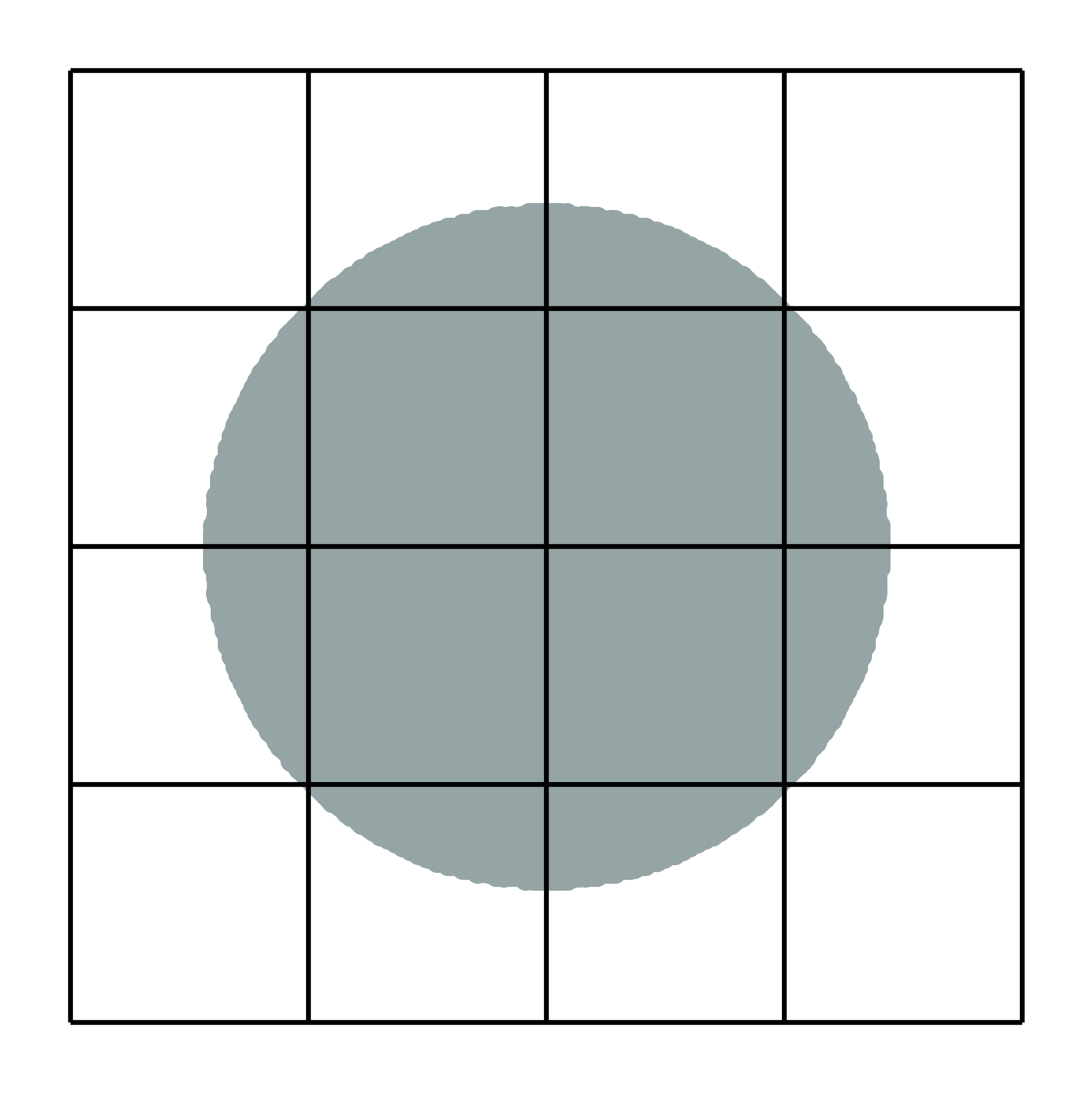}
    \includegraphics[width=0.4\columnwidth]{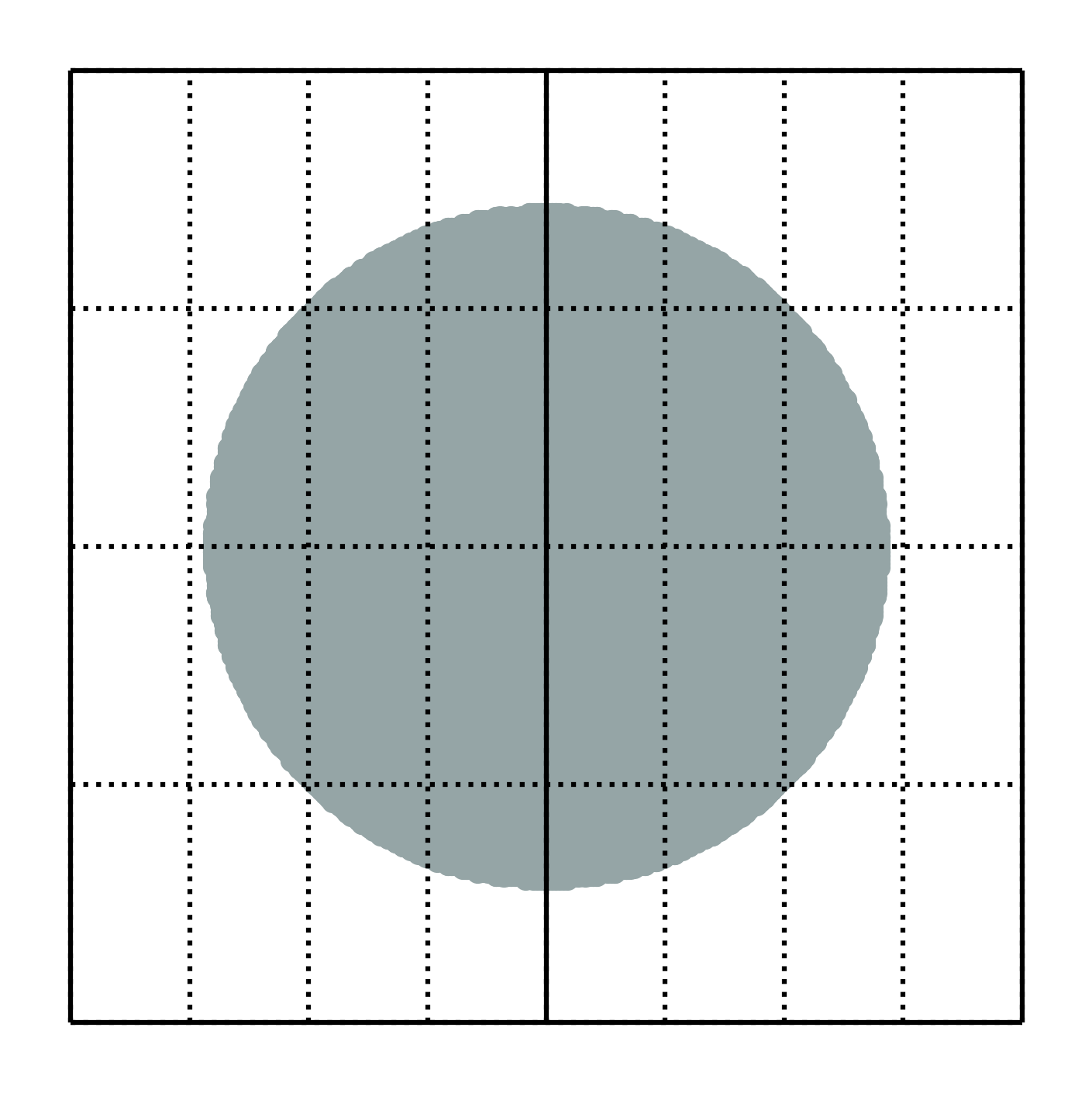}
    \caption{Initial spherical configuration of Lennard-Jones particles showing the domain decomposition at the $x$-$y$ plane with partitioning according to MPI ranks (---) and HPX subnodes ($\cdots$). Left: MPI-only partitioning into $(4,4,2)$ grid. Right: HPX-enabled partitioning into 2 MPI ranks and $(4,4,4)$ subnode grid.}
    \label{fig:mpi-vs-hpx-sphere-lj-setup}
\end{figure}

\begin{figure}[t!]
    \centering
    \begin{subfigure}[t]{0.6\columnwidth}
        \includegraphics[width=1.0\textwidth]{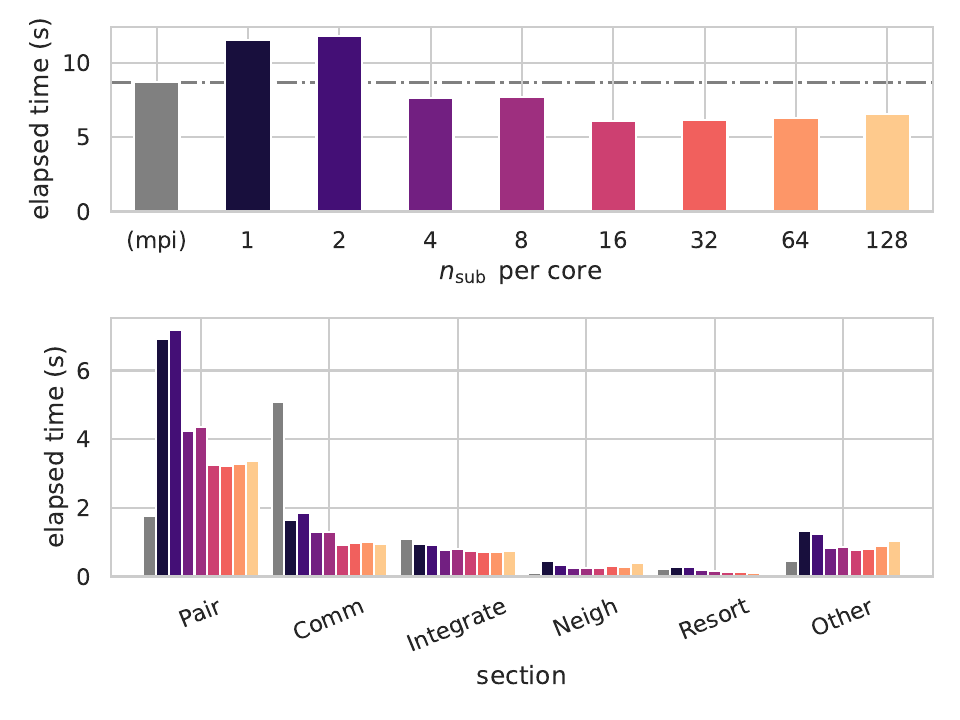}
        \caption{Mogon at 16 subnodes per core}
    \end{subfigure}
    \begin{subfigure}[t]{0.6\columnwidth}
        \includegraphics[width=1.0\textwidth]{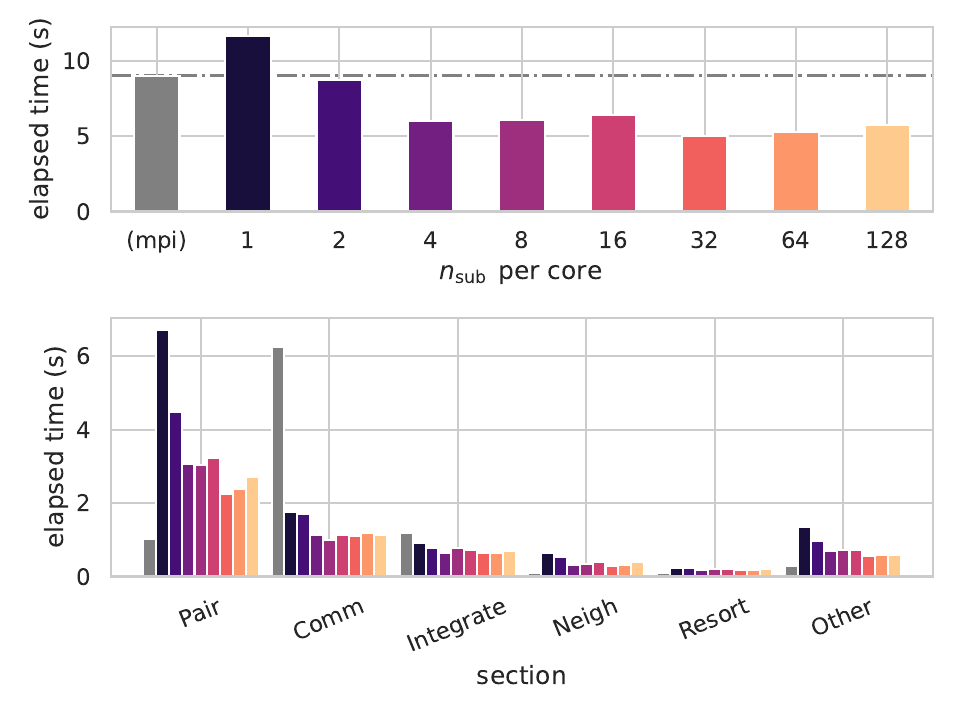}
        \caption{AMD at 32 subnodes per core}
    \end{subfigure}
    \caption{Comparison of sectional elapsed times with the MPI version ({\color{gray}-$\cdot$-$\cdot$}) and with HPX using different number of subnodes for the spherical Lennard-Jones. Lower time is better.}
    \label{fig:mpi-vs-hpx-sphere-lj}
\end{figure}

To investigate a spatially inhomogeneous case, we construct a simulation using Lennard-Jones particles that mimics  the load distribution arising in adaptive resolution simulations where particles in a spherical region are treated with full atomistic resolution, while the particles in the rest of the domain are coarse-grained \citep{fogarty2015adaptive} or treated as ideal gas \citep{baptista2021density} resulting in a locally reduced computational load.
As shown in Figure \ref{fig:mpi-vs-hpx-sphere-lj-setup}, a spherical region in the center of a simulation box is filled with Lennard-Jones particles with density $\rho_\text{in}=0.8442$ similar to the bulk simulation described earlier.

For this setup, we use a larger cubic simulation box of length $L=271$ with a spherical region containing 2.58 million particles or 16\% of the volume of the simulation box.
To keep the spherical structure intact, the thermostat temperature was kept at $T=0.1$ and the simulation was run for  100 time steps.

Results of the autotuning procedure for a full compute node are shown in Figure \ref{fig:mpi-vs-hpx-sphere-lj}.
In contrast to the bulk simulation, the optimal number of subnodes for this setup on Mogon has shifted to $n_\text{sub} = 256$ or 16 subnodes per core with a $1.4\times$ speedup over the MPI version.
On the other hand, the optimal $n_\text{sub}$ for both AMD and Raven clusters is 32 subnodes per core which indicates the need for the autotuning procedure since the same setup could behave differently on different systems.

Among the different code sections, the largest speedup can be observed in the communication which captures the MPI synchronization and takes the load imbalance into account.
In the MPI version, the processors with fewer particles in their subdomain have to wait for other processors with more particles to complete their force calculations and synchronize via MPI.
In the version with HPX multithreading, the work load is divided symmetrically between the two MPI ranks and the subnodes are executed concurrently by HPX threads.

In order to put these speedups into perspective, we devise a performance model to estimate the minimum execution time that we can ideally obtain for our load imbalanced system.
The theoretical minimum execution time $\tau_s$ for each section $s$ is obtained when the workload is evenly distributed across all MPI ranks, which we can relate to the corresponding measured execution time $t_s$ in the MPI version.
We observe that the timers for the \textsc{Pair}, \textsc{Neigh}, and \textsc{Other} sections are not affected by load imbalances since they do not involve any inter-node communication.
Thus, for these sections, $\tau_s = t_s$ since the measured execution times $t_s$ are already averages across MPI ranks.

For the remaining sections which involve MPI communication within the timers, we estimate the ideal time by running a homogenous setup and scaling the result using the relative number of particles.
We start with a bulk LJ setup formed by filling the entire simulation box described in Figure \ref{fig:mpi-vs-hpx-sphere-lj-setup} with LJ particles and measuring $t_s^\text{bulk}$, the execution time of each section.
From this, we can estimate $t_s$ by relating each section's workload to the relative number of particles in the two simulations ($N_\text{bulk}$ and $N_\text{sphere}$).
Since \textsc{Integrate} scales with the total number of particles in the simulation box, its ideal execution time is given by $\tau_\textsc{Integrate} = t_{\textsc{Integrate}}^\text{bulk} \left( {N_\text{sphere}}/{N_\text{bulk}} \right)$.
On the other hand, the \textsc{Resort} and and \textsc{Comm} sections are dominated by inter-node communication that happens only along the planes of the node boundaries which are proportional to the surface area of the simulation box.
Since the length of the boundaries is $L=\left( N / \rho \right)^{1/3}$ for some constant number density $\rho$,
the surface area of the simulation box is given by $6 \times \left( N / \rho \right)^{1/3} \times \left( N / \rho \right)^{1/3}$ which means that the communication workload is proportional to $N^{2/3}$.
Using this, we can estimate the ideal execution time for these sections to be
$\tau_\text{s} = t_{\text{s}}^\text{bulk} \left( {N_\text{sphere}}/{N_\text{bulk}} \right)^{2/3}$ for $s\in\{ \textsc{Resort}, \textsc{Comm} \}$.
Thus, the total ideal execution time for the spherical setup is given by
\begin{equation}
    \begin{split}
    \tau =~ & t_\textsc{Pair} + t_\textsc{Neigh} + t_\textsc{Other} +  t_{\textsc{Integrate}}^\text{bulk} \left( \frac{N_\text{sphere}}{N_\text{bulk}} \right) \\
        & + \left(t_{\textsc{Comm}}^\text{bulk} + t_{\textsc{Resort}}^\text{bulk} \right) \left( \frac{N_\text{sphere}}{N_\text{bulk}} \right)^{2/3}
    \end{split}
    \label{eq:perf-model}
\end{equation}

We performed the bulk simulation on the three machines and used Equation \eqref{eq:perf-model} to calculate the ideal execution times shown in Table \ref{tab:perf-model}. Also shown are the actual execution times using the HPX and MPI implementations from Figure \ref{fig:mpi-vs-hpx-sphere-lj}.
We find that on Mogon, our HPX implementation is $1.71\times$ slower compared to the ideal, while the base MPI implementation is $2.45\times$ slower.

\begin{table}
    \caption{Ideal execution time $\tau$ calculated from \eqref{eq:perf-model} compared with the actual execution time with HPX and MPI.}
    \centering
    \begin{tabular}{l|c|cc|cc}
        \toprule
        & $\tau$ (s) & $t_\text{HPX}$ (s) & $t_\text{HPX}/\tau$ & $t_\text{MPI}$ (s) & $t_\text{MPI}/\tau$ \\
        \midrule
        Mogon & 3.55 & 6.07 & 1.71 & 8.69 & 2.45 \\
        Raven & 1.75 & 3.12 & 1.78 & 5.39 & 3.08 \\
        AMD & 2.34 & 5.05 & 2.16 & 8.99 & 3.84 \\
        \bottomrule
    \end{tabular}
    \label{tab:perf-model}
\end{table}

\section{Related work}
\label{sec:related-work}

Maximizing SIMD capabilities of recent hardware architectures has been thoroughly explored and applied to molecular dynamics simulations.
For example, optimizations have been made in LAMMPS that enable full use of the Intel Knights Landing architecture with AVX-512 which is still relevant to  recent Intel CPUs \citep{jeffers2016intel}.
SIMD vectorization techniques have also been investigated specifically for the Lennard-Jones potential on Intel CPUs with AVX2 and AVX-512 instructions \citep{watanabe2019simd}.

Different forms of the hybrid MPI+X programming model have been widely applied in molecular dynamics simulations.
For example, LAMMPS has support for multi-core CPUs and accelerators through OpenMP, CUDA and Kokkos \citep{jeffers2016intel, lammpsstable29Oct2020}.
To reduce the complexity of maintaining code for different architectures, performance portability has also been applied to particle-based simulations through Cabana, which is a library built on Kokkos, and demonstrated specifically for molecular dynamics simulations through the CabanaMD proxy application \citep{mniszewski2021enabling}.
Task-based parallelism and asynchronous mesh refinement approaches have previously been implemented for highly dynamic MD simulations parallelized using MPI and OpenMP \citep{prat2020amr}.
It has also been shown that fully adopting a hybrid MPI+OpenMP programming model alleviates load balancing problems and communication pressure better than pure MPI implementation that allows the adjustment of domain boundaries \citep{morillo2022hybrid}.

HPX has been successfully applied to N-body simulations of stellar mergers in astrophysics \citep{marcello2021octo}.
The implication of task granularity on performance of HPX-based applications have also been studied \citep{grubel2015performance}.
Comparisons between the traditional MPI programming model and HPX-based implementations of the discontinuous Galerkin method for the two-dimensional shallow water equations have also been reported in \citep{bremer2019performance} which found a 6\% faster runtime with HPX parallelization which they attribute to better cache behavior due to overdecomposition.

\section{Summary and discussion}
\label{sec:summary-discussion}

In this work, we presented our code modernization efforts for the ESPResSo++ molecular dynamics simulation package.
This was achieved by optimizing the particle data layout, implementing SIMD vectorization on critical loops and integrating node-level parallelism through HPX.
We have optimized operations that require only a few particle attributes by using a structure of arrays layout which improved the memory access patterns of these operations resulting in an up to $2\times$ overall speedup.
The change in layout also allowed us to ensure SIMD vectorization on critical for loops used in short-range interactions. This results in an additional $1.5\times$ speedup for Lennard-Jones simulations and $1.1\times$ speedup for polymer melt simulations, suggesting that the greater benefits from vectorization of short-range interactions depend on having a larger cut-off distance among other factors.

These serial optimizations served as a baseline on which we implemented node-level parallelism through HPX.
By decomposing the MPI node into smaller subnodes, we were able to modulate the granularity of the task sizes and obtain an optimum balance between resource starvation and overhead from additional force calculation between subnodes.
We then compared the HPX-based implementation with the traditional MPI-based parallelization.
For a spatially homogenous system, we found the HPX implementation to be $5\%$ slower on a single socket running a single MPI rank, which rises to $37\%$ slower on a full node running two MPI ranks resulting from the increase in communication volume per MPI call.
Nevertheless, the benefit of using the work-stealing capabilities of HPX was demonstrated when running spatially inhomogeneous simulations such as an LJ system with spherical load distribution for which we obtained a $1.4\times$ speedup.

To maintain compatibility with the original packages within ESPResSo++, we have chosen to implement our code modernization strategies as additional submodules in ESPResSo++ (\texttt{vec} and \texttt{hpx4espp}).
This is due to the intrusive change in the particle data structure which all algorithms and analysis tools are dependent on.
This design choice allows us to continuously develop other performance improvements which could eventually be adopted in the main branch of ESPResSo++.

Through this implementation and evaluation process, we have found that HPX can provide suitable features for implementing thread-level parallelism for MD applications involving short-range interactions while keeping MPI as the driver for inter-node communication.
However, as with the introduction of any shared-memory programming model, certain care has to be taken in order to ensure thread-safety, which is not a concern in MPI-based implementations.
We were able to solve this in our HPX implementation by introducing a node-level domain decomposition layer at the expense of
some additional overhead.
By autotuning for a few time steps, we can find an optimal task size that maximizes parallelism and minimizes this overhead.

\section{Future work}
\label{sec:future-work}

Further performance improvements can be achieved by utilizing HPX also for the distributed parallelism aspect of our application.
The current implementation, which relies on the multithreaded parallel algorithms of HPX, is still limited by implicit synchronization among the different MPI localities since they still utilize MPI communication to update their ghost cells with only the main thread handling MPI calls.
This constraint can be further decoupled by allowing every subnode to independently perform ghost cell updates using asynchronous remote function calls.
This can be done using the distributed parallelism framework of HPX through its Active Global Address Space (AGAS).
Using global addressing, we can decouple the ghost communication among neighboring subnodes so that every direction can be updated independently through an API that is the same whether the neighboring subnode is in the same locality or not.
This can then utilize the futurization capabilities of HPX to form a task-based dependency tree that will expose more parallelism and reduce the synchronization among MPI localities.

Although this work only addresses node-level load balancing through work-stealing, HPX can also be used to implement load balancing among different localities by converting the subnodes into migratable objects that can be moved between localities while maintaining their neighborhood topologies through their global addressability.

\section*{Declaration}
The authors declare that they have no known competing financial interests or personal relationships that could have appeared to influence the work reported in this paper.

\section*{Funding}
This work was funded by the German Research Foundation (DFG) through the collaborative research center TRR 146 (Project G).
ESPResSo++ is one of the central software packages of the TRR146 (\url{https://trr146.de}) and also part of the software pool of the European E-CAM project (\url{https://www.e-cam2020.eu}).
Parts of this research were conducted using the supercomputer MOGON II and advisory services offered by Johannes Gutenberg University Mainz (\url{https://hpc.uni-mainz.de}), which is a member of the AHRP (Alliance for High Performance Computing in Rhineland Palatinate, \url{https://www.ahrp.info}) and the Gauss Alliance.
Some of the development was also carried out with the HPC system Raven at the Max Planck Computing and Data Facility (\url{https://www.mpcdf.mpg.de}).
LANL is operated by Triad National Security, LLC, for the National Nuclear Security Administration of U.S. Department of Energy (Contract No. 89233218CNA000001). This document is LA-UR-22-21943.

\bibliography{readcube.bib,references.bib}

\begin{thebibliography}{10}
\expandafter\ifx\csname url\endcsname\relax
  \def\url#1{\texttt{#1}}\fi
\expandafter\ifx\csname urlprefix\endcsname\relax\def\urlprefix{URL }\fi
\expandafter\ifx\csname href\endcsname\relax
  \def\href#1#2{#2} \def\path#1{#1}\fi

\bibitem{plimpton1995fast}
S.~Plimpton, {Fast Parallel Algorithms for Short-Range Molecular Dynamics},
  Journal of Computational Physics 117~(1) (1995) 1--19.
\newblock \href {https://doi.org/10.1006/jcph.1995.1039}
  {\path{doi:10.1006/jcph.1995.1039}}.

\bibitem{thompson2022lammps}
A.~P. Thompson, H.~M. Aktulga, R.~Berger, D.~S. Bolintineanu, W.~M. Brown,
  P.~S. Crozier, P.~J. in't Veld, A.~Kohlmeyer, S.~G. Moore, T.~D. Nguyen,
  et~al., {LAMMPS-a flexible simulation tool for particle-based materials
  modeling at the atomic, meso, and continuum scales}, Computer Physics
  Communications 271 (2022) 108171.
\newblock \href {https://doi.org/10.1016/j.cpc.2021.108171}
  {\path{doi:10.1016/j.cpc.2021.108171}}.

\bibitem{abraham2015gromacs}
M.~J. Abraham, T.~Murtola, R.~Schulz, S.~Páll, J.~C. Smith, B.~Hess,
  E.~Lindahl, {GROMACS: High performance molecular simulations through
  multi-level parallelism from laptops to supercomputers}, SoftwareX 1 (2015)
  19--25.
\newblock \href {https://doi.org/10.1016/j.softx.2015.06.001}
  {\path{doi:10.1016/j.softx.2015.06.001}}.

\bibitem{phillips2020scalable}
J.~C. Phillips, D.~J. Hardy, J.~D.~C. Maia, J.~E. Stone, J.~V. Ribeiro, R.~C.
  Bernardi, R.~Buch, G.~Fiorin, J.~Hénin, W.~Jiang, R.~McGreevy, M.~C.~R.
  Melo, B.~K. Radak, R.~D. Skeel, A.~Singharoy, Y.~Wang, B.~Roux,
  A.~Aksimentiev, Z.~Luthey-Schulten, L.~V. Kalé, K.~Schulten, C.~Chipot,
  E.~Tajkhorshid, {Scalable molecular dynamics on CPU and GPU architectures
  with NAMD}, The Journal of Chemical Physics 153~(4) (2020) 044130.
\newblock \href {https://doi.org/10.1063/5.0014475}
  {\path{doi:10.1063/5.0014475}}.

\bibitem{halverson2013espressopp}
J.~D. Halverson, T.~Brandes, O.~Lenz, A.~Arnold, S.~Bevc, V.~Starchenko,
  K.~Kremer, T.~Stuehn, D.~Reith, {ESPResSo++: A modern multiscale simulation
  package for soft matter systems}, Computer Physics Communications 184~(4)
  (2013) 1129--1149.
\newblock \href {https://doi.org/10.1016/j.cpc.2012.12.004}
  {\path{doi:10.1016/j.cpc.2012.12.004}}.

\bibitem{guzman2019espresso++}
H.~V. Guzman, N.~Tretyakov, H.~Kobayashi, A.~C. Fogarty, K.~Kreis, J.~Krajniak,
  C.~Junghans, K.~Kremer, T.~Stuehn, {ESPResSo++ 2.0: Advanced methods for
  multiscale molecular simulation}, Computer Physics Communications 238 (2019)
  66--76.
\newblock \href {http://arxiv.org/abs/1806.10841} {\path{arXiv:1806.10841}},
  \href {https://doi.org/10.1016/j.cpc.2018.12.017}
  {\path{doi:10.1016/j.cpc.2018.12.017}}.

\bibitem{watanabe2019simd}
H.~Watanabe, K.~M. Nakagawa, {SIMD vectorization for the Lennard-Jones
  potential with AVX2 and AVX-512 instructions}, Computer Physics
  Communications 237 (2019) 1--7.
\newblock \href {http://arxiv.org/abs/1806.05713} {\path{arXiv:1806.05713}},
  \href {https://doi.org/10.1016/j.cpc.2018.10.028}
  {\path{doi:10.1016/j.cpc.2018.10.028}}.

\bibitem{kretz2012vc}
M.~Kretz, V.~Lindenstruth, {Vc: A C++ library for explicit vectorization},
  Software: Practice and Experience 42~(11) (2012) 1409--1430.
\newblock \href {https://doi.org/10.1002/spe.1149}
  {\path{doi:10.1002/spe.1149}}.

\bibitem{ishiyama2012fflops}
T.~Ishiyama, K.~Nitadori, J.~Makino, 4.45 pflops astrophysical n-body
  simulation on k computer -- the gravitational trillion-body problem, in: SC
  '12: Proceedings of the International Conference on High Performance
  Computing, Networking, Storage and Analysis, 2012, pp. 1--10.
\newblock \href {https://doi.org/10.1109/SC.2012.3}
  {\path{doi:10.1109/SC.2012.3}}.

\bibitem{fattebert2012dynamic}
J.-L. Fattebert, D.~Richards, J.~Glosli,
  \href{https://www.sciencedirect.com/science/article/pii/S0010465512002524}{Dynamic
  load balancing algorithm for molecular dynamics based on voronoi cells domain
  decompositions}, Computer Physics Communications 183~(12) (2012) 2608--2615.
\newblock \href {https://doi.org/https://doi.org/10.1016/j.cpc.2012.07.013}
  {\path{doi:https://doi.org/10.1016/j.cpc.2012.07.013}}.
\newline\urlprefix\url{https://www.sciencedirect.com/science/article/pii/S0010465512002524}

\bibitem{Eibl2019}
S.~Eibl, U.~Rüde, A systematic comparison of runtime load balancing algorithms
  for massively parallel rigid particle dynamics, Computer Physics
  Communications 244 (2019) 76--85.
\newblock \href {https://doi.org/10.1016/j.cpc.2019.06.020}
  {\path{doi:10.1016/j.cpc.2019.06.020}}.

\bibitem{RaicuFZ08}
I.~Raicu, I.~T. Foster, Y.~Zhao, Many-task computing for grids and
  supercomputers, in: 2008 Workshop on Many-Task Computing on Grids and
  Supercomputers (MTAGS@SC), Austin, TX, USA, November 17, 2008, pp. 1--11.
\newblock \href {https://doi.org/10.1109/MTAGS.2008.4777912}
  {\path{doi:10.1109/MTAGS.2008.4777912}}.

\bibitem{kale2020charm++short}
L.~Kale, et~al., Uiuc-ppl/charm: v7.0.0-rc1 (Jun. 2021).
\newblock \href {https://doi.org/10.5281/zenodo.4988098}
  {\path{doi:10.5281/zenodo.4988098}}.

\bibitem{bakosi2021asynchronous}
J.~Bakosi, R.~Bird, F.~Gonzalez, C.~Junghans, W.~Li, H.~Luo, A.~Pandare,
  J.~Waltz, Asynchronous distributed-memory task-parallel algorithm for
  compressible flows on unstructured 3d eulerian grids, Advances in Engineering
  Software 160 (2021) 102962.
\newblock \href {https://doi.org/10.1016/j.advengsoft.2020.102962}
  {\path{doi:10.1016/j.advengsoft.2020.102962}}.

\bibitem{kaiser2020hpx}
H.~Kaiser, P.~Diehl, A.~Lemoine, B.~Lelbach, P.~Amini, A.~Berge,
  J.~Biddiscombe, S.~Brandt, N.~Gupta, T.~Heller, K.~Huck, Z.~Khatami,
  A.~Kheirkhahan, A.~Reverdell, S.~Shirzad, M.~Simberg, B.~Wagle, W.~Wei,
  T.~Zhang, {HPX - The C++ Standard Library for Parallelism and Concurrency},
  Journal of Open Source Software 5~(53) (2020) 2352.
\newblock \href {https://doi.org/10.21105/joss.02352}
  {\path{doi:10.21105/joss.02352}}.

\bibitem{allen2017computer}
M.~P. Allen, D.~J. Tildesley, {Computer Simulation of Liquids}, {Oxford
  university press}, 2017.
\newblock \href {https://doi.org/10.1093/oso/9780198803195.001.0001}
  {\path{doi:10.1093/oso/9780198803195.001.0001}}.

\bibitem{verlet1967computer}
L.~Verlet, {Computer "Experiments" on Classical Fluids. I. Thermodynamical
  Properties of Lennard-Jones Molecules}, Physical Review 159~(1) (1967)
  98--103.
\newblock \href {https://doi.org/10.1103/physrev.159.98}
  {\path{doi:10.1103/physrev.159.98}}.

\bibitem{swope1982computer}
W.~C. Swope, H.~C. Andersen, P.~H. Berens, K.~R. Wilson, {A computer simulation
  method for the calculation of equilibrium constants for the formation of
  physical clusters of molecules: Application to small water clusters}, The
  Journal of Chemical Physics 76~(1) (1982) 637--649.
\newblock \href {https://doi.org/10.1063/1.442716}
  {\path{doi:10.1063/1.442716}}.

\bibitem{ruhle2009versatile}
V.~Rühle, C.~Junghans, A.~Lukyanov, K.~Kremer, D.~Andrienko, {Versatile
  Object-Oriented Toolkit for Coarse-Graining Applications}, Journal of
  Chemical Theory and Computation 5~(12) (2009) 3211--3223.
\newblock \href {https://doi.org/10.1021/ct900369w}
  {\path{doi:10.1021/ct900369w}}.

\bibitem{wehner2018electronic}
J.~Wehner, L.~Brombacher, J.~Brown, C.~Junghans, O.~{\c{C}}aylak, Y.~Khalak,
  P.~Madhikar, G.~Tirimb{\`o}, B.~Baumeier, Electronic excitations in complex
  molecular environments: Many-body green’s functions theory in votca-xtp,
  Journal of chemical theory and computation 14~(12) (2018) 6253--6268.
\newblock \href {https://doi.org/10.1021/acs.jctc.8b00617}
  {\path{doi:10.1021/acs.jctc.8b00617}}.

\bibitem{xu2021implementation}
Z.-H. Xu, J.~Vance, N.~Tretyakov, T.~Stuehn, A.~Brinkmann, {Implementation and
  Code Parallelization of the Lees-Edwards Boundary Condition in ESPResSo++},
  arXiv preprint (2021).
\newblock \href {https://doi.org/10.48550/arXiv.2109.11083}
  {\path{doi:10.48550/arXiv.2109.11083}}.

\bibitem{wolfe2014compilers}
M.~Wolfe, \href{https://www.hpcwire.com/2014/07/16/compilers-mpix}{{Compilers
  and more: MPI+ X}}, HPC Wire (2014).
\newline\urlprefix\url{https://www.hpcwire.com/2014/07/16/compilers-mpix}

\bibitem{trott2022Kokkos}
C.~R. Trott, D.~Lebrun-Grandié, D.~Arndt, J.~Ciesko, V.~Dang, N.~Ellingwood,
  R.~Gayatri, E.~Harvey, D.~S. Hollman, D.~Ibanez, N.~Liber, J.~Madsen,
  J.~Miles, D.~Poliakoff, A.~Powell, S.~Rajamanickam, M.~Simberg,
  D.~Sunderland, B.~Turcksin, J.~Wilke, {Kokkos 3: Programming Model Extensions
  for the Exascale Era}, IEEE Transactions on Parallel and Distributed Systems
  33~(4) (2022) 805--817.
\newblock \href {https://doi.org/10.1109/TPDS.2021.3097283}
  {\path{doi:10.1109/TPDS.2021.3097283}}.

\bibitem{beckingsale2019raja}
D.~A. Beckingsale, J.~Burmark, R.~Hornung, H.~Jones, W.~Killian, A.~J. Kunen,
  O.~Pearce, P.~Robinson, B.~S. Ryujin, T.~R. Scogland, {RAJA: Portable
  performance for large-scale scientific applications}, in: 2019 ieee/acm
  international workshop on performance, portability and productivity in hpc
  (p3hpc), IEEE, 2019, pp. 71--81.
\newblock \href {https://doi.org/10.1109/P3HPC49587.2019.00012}
  {\path{doi:10.1109/P3HPC49587.2019.00012}}.

\bibitem{gutierrez2017accommodating}
S.~K. Guti{\'e}rrez, K.~Davis, D.~C. Arnold, R.~S. Baker, R.~W. Robey,
  P.~McCormick, D.~Holladay, J.~A. Dahl, R.~J. Zerr, F.~Weik, et~al.,
  Accommodating thread-level heterogeneity in coupled parallel applications,
  in: 2017 IEEE International Parallel and Distributed Processing Symposium
  (IPDPS), IEEE, 2017, pp. 469--478.
\newblock \href {https://doi.org/10.1109/IPDPS.2017.13}
  {\path{doi:10.1109/IPDPS.2017.13}}.

\bibitem{jeffers2016intel}
J.~Jeffers, J.~Reinders, A.~Sodani,
  \href{https://dl.acm.org/doi/10.5555/3050856}{{Intel Xeon Phi processor high
  performance programming: knights landing edition}}, Morgan Kaufmann, 2016.
\newline\urlprefix\url{https://dl.acm.org/doi/10.5555/3050856}

\bibitem{bremer2019performance}
M.~Bremer, K.~Kazhyken, H.~Kaiser, C.~Michoski, C.~Dawson, {Performance
  Comparison of HPX Versus Traditional Parallelization Strategies for the
  Discontinuous Galerkin Method}, Journal of Scientific Computing 80~(2) (2019)
  878--902.
\newblock \href {https://doi.org/10.1007/s10915-019-00960-z}
  {\path{doi:10.1007/s10915-019-00960-z}}.

\bibitem{grubel2015performance}
P.~Grubel, H.~Kaiser, J.~Cook, A.~Serio,
  \href{http://stellar.cct.lsu.edu/pubs/hpcmaspa2015.pdf}{{The Performance
  Implication of Task Size for Applications on the HPX Runtime System}}, 2015
  IEEE International Conference on Cluster Computing (2015) 682--689\href
  {https://doi.org/10.1109/cluster.2015.119}
  {\path{doi:10.1109/cluster.2015.119}}.
\newline\urlprefix\url{http://stellar.cct.lsu.edu/pubs/hpcmaspa2015.pdf}

\bibitem{kaiser2020hpx151short}
H.~Kaiser, et~al.,
  \href{https://doi.org/10.5281/zenodo.4059746}{{STEllAR-GROUP/hpx: HPX V1.5.1:
  The C++ Standards Library for Parallelism and Concurrency}} (Sep. 2020).
\newblock \href {https://doi.org/10.5281/zenodo.4059746}
  {\path{doi:10.5281/zenodo.4059746}}.
\newline\urlprefix\url{https://doi.org/10.5281/zenodo.4059746}

\bibitem{kremer1990dynamics}
K.~Kremer, G.~S. Grest, {Dynamics of entangled linear polymer melts: A
  molecular‐dynamics simulation}, The Journal of Chemical Physics 92~(8)
  (1990) 5057--5086.
\newblock \href {https://doi.org/10.1063/1.458541}
  {\path{doi:10.1063/1.458541}}.

\bibitem{lammpsstable29Oct2020}
S.~Plimpton, A.~Kohlmeyer, A.~Thompson, S.~Moore, R.~Berger, {LAMMPS Stable
  release 29 October 2020} (Oct. 2020).
\newblock \href {https://doi.org/10.5281/zenodo.4157471}
  {\path{doi:10.5281/zenodo.4157471}}.

\bibitem{fogarty2015adaptive}
A.~C. Fogarty, R.~Potestio, K.~Kremer, Adaptive resolution simulation of a
  biomolecule and its hydration shell: structural and dynamical properties, The
  Journal of Chemical Physics 142~(19) (2015) 05B610\_1.
\newblock \href {https://doi.org/10.1063/1.4921347}
  {\path{doi:10.1063/1.4921347}}.

\bibitem{baptista2021density}
L.~A. Baptista, R.~C. Dutta, M.~Sevilla, M.~Heidari, R.~Potestio, K.~Kremer,
  R.~Cortes-Huerto, Density-functional-theory approach to the hamiltonian
  adaptive resolution simulation method, Journal of Physics: Condensed Matter
  33~(18) (2021) 184003.
\newblock \href {https://doi.org/10.1088/1361-648X/abed1d}
  {\path{doi:10.1088/1361-648X/abed1d}}.

\bibitem{mniszewski2021enabling}
S.~M. Mniszewski, J.~Belak, J.-L. Fattebert, C.~F. Negre, S.~R. Slattery, A.~A.
  Adedoyin, R.~F. Bird, C.~Chang, G.~Chen, S.~Ethier, et~al., {Enabling
  particle applications for exascale computing platforms}, The International
  Journal of High Performance Computing Applications 35~(6) (2021) 572--597.
\newblock \href {https://doi.org/10.1177/10943420211022829}
  {\path{doi:10.1177/10943420211022829}}.

\bibitem{prat2020amr}
R.~Prat, T.~Carrard, L.~Soulard, O.~Durand, R.~Namyst, L.~Colombet, {AMR-based
  molecular dynamics for non-uniform, highly dynamic particle simulations},
  Computer Physics Communications 253 (2020) 107177.
\newblock \href {https://doi.org/10.1016/j.cpc.2020.107177}
  {\path{doi:10.1016/j.cpc.2020.107177}}.

\bibitem{morillo2022hybrid}
J.~Morillo, M.~Vassaux, P.~V. Coveney, M.~Garcia-Gasulla, Hybrid
  parallelization of molecular dynamics simulations to reduce load imbalance,
  The Journal of Supercomputing (2022) 1--32\href
  {https://doi.org/10.1007/s11227-021-04214-4}
  {\path{doi:10.1007/s11227-021-04214-4}}.

\bibitem{marcello2021octo}
D.~C. Marcello, S.~Shiber, O.~De~Marco, J.~Frank, G.~C. Clayton, P.~M. Motl,
  P.~Diehl, H.~Kaiser, {OCTO-TIGER: a new, 3D hydrodynamic code for stellar
  mergers that uses HPX parallelization}, Monthly Notices of the Royal
  Astronomical Society 504~(4) (2021) 5345--5382.
\newblock \href {https://doi.org/10.1093/mnras/stab937}
  {\path{doi:10.1093/mnras/stab937}}.

\end{thebibliography}

\end{document}


\begin{frontmatter}
		
		\title{Supplementary Information to "Code modernization strategies for short-range non-bonded molecular dynamics simulations"}
		
		\author[affil1]{James Vance\corref{mycorrespondingauthor}}
		\ead{jnvance@outlook.com}
		
		\author[affil1]{Zhen-Hao Xu}
		\author[affil1]{Nikita Tretyakov}
		\author[affil2]{Torsten Stuehn}
		\author[affil3]{Markus Rampp}
		\author[affil3]{Sebastian Eibl}
		\author[affil4]{Christoph Junghans}
		\author[affil1]{Andr\'e Brinkmann\corref{mycorrespondingauthor}}
		\cortext[mycorrespondingauthor]{Corresponding author}
		\ead{brinkman@uni-mainz.de}
		
		\address[affil1]{Zentrum f\"ur Datenverarbeitung, Johannes Gutenberg-Universit\"at Mainz, Mainz, Germany}
		\address[affil2]{Max Planck Institute for Polymer Research, Mainz, Germany}
		\address[affil3]{Max Planck Computing and Data Facility, Garching, Germany}
		\address[affil4]{Applied Computer Science Group, Los Alamos National Laboratory, Los Alamos, New Mexico, USA}
		
	\end{frontmatter}
	
	\begin{figure}
		\centering
		a)
		\includegraphics[width=0.75\textwidth]{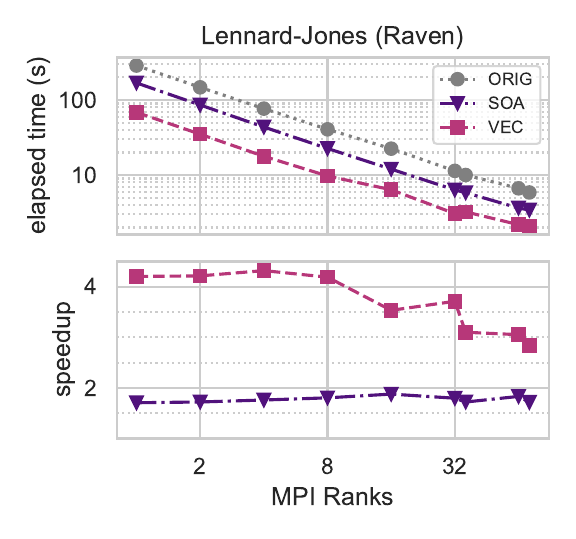}
		
		b)
		\includegraphics[width=0.75\textwidth]{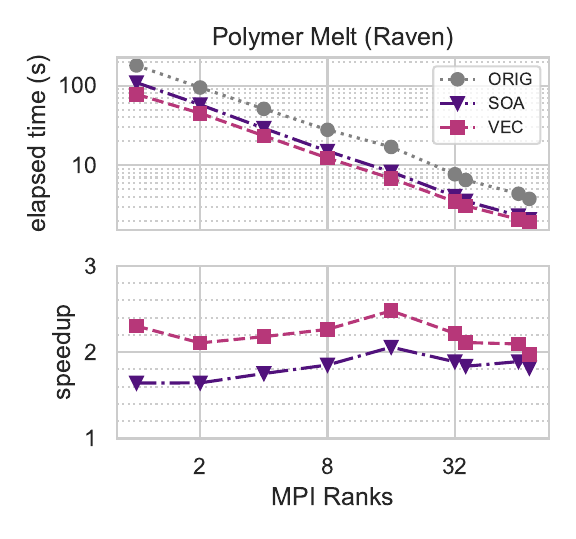}
	\end{figure}
	\begin{figure}
		\centering
		c) 
		\includegraphics[width=0.75\textwidth]{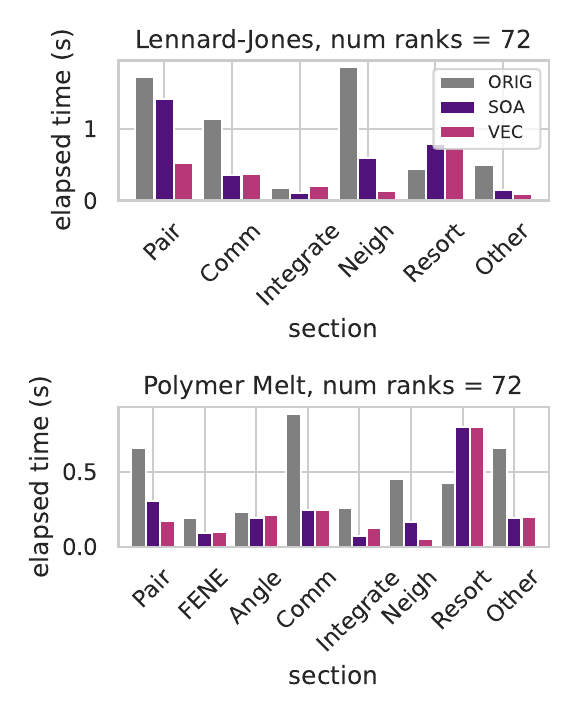}
		\label{subfig:soa-vec-sec-raven}
		
		\caption{Elapsed time and speedups for simulation in a) LJ fluid and b) polymer melts on one \textbf{Raven} computer node, including c) the sectional contributions. This figure is supplementary to Fig. 5 in the main text.}
		\label{fig-si:soa-vec-run-raven}
	\end{figure}
	
	\clearpage
	\begin{table}[h!]
		\caption{Raw data of elapsed runtimes (s) and speedups to simulation in LJ fluid and polymer melts on \textbf{Mogon}.}
		\centering
		\resizebox{\textwidth}{!}{
		\begin{tabular}{c |r r r |r r|r r r |r r}
			\toprule
			\multirow{3}{*}{\tabincell{c}{MPI\\ranks}} & \multicolumn{5}{c|}{Lennard-Jones fluid} &
			\multicolumn{5}{c}{polymer melts} \\
			\cline{2-11}
			& \multicolumn{3}{c|}{elapsed time} & \multicolumn{2}{c|}{speedup}& \multicolumn{3}{c|}{elapsed time} & \multicolumn{2}{c}{speedup}\\
			\cline{2-11}
			 & \textsc{orig} & \textsc{soa} & \textsc{vec} & \textsc{soa} & \textsc{vec}& \textsc{orig} & \textsc{soa} & \textsc{vec} & \textsc{soa} & \textsc{vec} \\
			\midrule
1 & 230.0 & 126.4 & 61.6 & 1.82 & 3.74 & 198.5 & 116.4 & 105.7 & 1.71 & 1.88 \\
2 & 116.0 & 64.0 & 31.9 & 1.81 & 3.64 & 97.6 & 51.5 & 45.9 & 1.90 & 2.13     \\
4 & 63.4 & 34.5 & 17.5 & 1.83 & 3.62 & 52.6 & 28.4 & 25.6 & 1.85 & 2.05      \\
8 & 33.9 & 19.1 & 10.6 & 1.77 & 3.18 & 28.1 & 15.3 & 13.5 & 1.84 & 2.08      \\
16 & 25.9 & 12.7 & 8.0 & 2.04 & 3.25 & 21.1 & 12.1 & 11.2 & 1.75 & 1.88      \\
32 & 13.3 & 6.8 & 4.4 & 1.97 & 3.05 & 10.4 & 5.3 & 4.8 & 1.97 & 2.15         \\
			\bottomrule
		\end{tabular}
	}
	\end{table}
\begin{table}[h!]
\caption{Raw data of sectional runtimes to simulation in LJ fluid and polymer melts on \textbf{Mogon}.}
\centering
\begin{tabular}{l |ccc|ccc }
	\toprule
	\multirow{2}{*}{Section} & \multicolumn{3}{c|}{Lennard-Jones fluid} &
	\multicolumn{3}{c}{polymer melts} \\
	\cline{2-7}
	& \textsc{orig} & \textsc{soa} & \textsc{vec}& \textsc{orig} & \textsc{soa} & \textsc{vec}\\
\midrule
Pair & 3.87 & 3.31 & 1.81 & 1.41 & 0.81 & 0.59       \\
FENE & / & / & / & 0.77 & 0.39 & 0.41                \\
Angle & / & / & / & 1.21 & 0.61 & 0.69               \\
Comm & 2.23 & 0.50 & 0.45 & 1.54 & 0.55 & 0.52       \\
Integrate & 0.80 & 0.16 & 0.19 & 1.62 & 0.43 & 0.41  \\
Neigh & 4.08 & 1.20 & 0.39 & 1.02 & 0.37 & 0.19      \\
Resort & 0.86 & 1.26 & 1.26 & 1.00 & 1.63 & 1.61     \\
Other & 1.49 & 0.35 & 0.26 & 1.83 & 0.48 & 0.41      \\
	\bottomrule
\end{tabular}
\end{table}
	
\clearpage
\begin{table}[h!]
	\caption{Raw data of elapsed runtimes (s) and speedups to simulation in LJ fluid and polymer melts on \textbf{AMD}.}
	\centering
	\resizebox{\textwidth}{!}{
		\begin{tabular}{c |r r r |r r|r r r |r r}
			\toprule
			\multirow{3}{*}{\tabincell{c}{MPI\\ranks}} & \multicolumn{5}{c|}{Lennard-Jones fluid} &
			\multicolumn{5}{c}{polymer melts} \\
			\cline{2-11}
			& \multicolumn{3}{c|}{elapsed time} & \multicolumn{2}{c|}{speedup}& \multicolumn{3}{c|}{elapsed time} & \multicolumn{2}{c}{speedup}\\
			\cline{2-11}
			& \textsc{orig} & \textsc{soa} & \textsc{vec} & \textsc{soa} & \textsc{vec}& \textsc{orig} & \textsc{soa} & \textsc{vec} & \textsc{soa} & \textsc{vec} \\
			\midrule
1 & 260.5 & 171.5 & 151.3 & 1.52 & 1.72 & 188.5 & 133.6 & 120.2 & 1.41 & 1.57 \\
2 & 138.0 & 87.5 & 78.1 & 1.58 & 1.77 & 105.0 & 75.4 & 68.3 & 1.39 & 1.54     \\
4 & 73.3 & 45.2 & 40.6 & 1.62 & 1.81 & 56.5 & 39.0 & 35.7 & 1.45 & 1.58       \\
8 & 37.5 & 23.2 & 20.8 & 1.62 & 1.81 & 28.2 & 15.7 & 14.5 & 1.80 & 1.95       \\
16 & 19.1 & 12.5 & 11.2 & 1.53 & 1.70 & 13.0 & 8.2 & 7.7 & 1.59 & 1.68        \\
32 & 9.6 & 6.6 & 6.0 & 1.44 & 1.59 & 5.1 & 4.5 & 4.2 & 1.13 & 1.20            \\
64 & 6.8 & 4.6 & 4.2 & 1.50 & 1.64 & 3.5 & 3.2 & 3.1 & 1.09 & 1.13            \\
			\bottomrule
		\end{tabular}
	}
\end{table}
\begin{table}[h!]
	\caption{Raw data of sectional runtimes to simulation in LJ fluid and polymer melts on \textbf{AMD}.}
	\centering
	\begin{tabular}{l |ccc|ccc }
		\toprule
		\multirow{2}{*}{Section} & \multicolumn{3}{c|}{Lennard-Jones fluid} &
		\multicolumn{3}{c}{polymer melts} \\
		\cline{2-7}
		& \textsc{orig} & \textsc{soa} & \textsc{vec}& \textsc{orig} & \textsc{soa} & \textsc{vec}\\
		\midrule
Pair & 2.88 & 2.15 & 1.79 & 0.62 & 0.69 & 0.37      \\
FENE & / & / & / & 0.19 & 0.11 & 0.20               \\
Angle & / & / & / & 0.36 & 0.29 & 0.42              \\
Comm & 1.47 & 0.73 & 0.69 & 0.81 & 0.63 & 0.62      \\
Integrate & 0.26 & 0.08 & 0.08 & 0.20 & 0.17 & 0.16 \\
Neigh & 1.43 & 0.66 & 0.67 & 0.33 & 0.17 & 0.18     \\
Resort & 0.45 & 0.79 & 0.78 & 0.52 & 0.90 & 0.91    \\
Other & 0.35 & 0.16 & 0.16 & 0.42 & 0.22 & 0.21     \\
		\bottomrule
	\end{tabular}
\end{table}
\clearpage
\begin{table}[h!]
\caption{Raw data of elapsed runtimes (s) and speedups to simulation in LJ fluid and polymer melts on \textbf{Raven}.}
\centering
\resizebox{\textwidth}{!}{
	\begin{tabular}{c |r r r |r r|r r r |r r}
		\toprule
		\multirow{3}{*}{\tabincell{c}{MPI\\ranks}} & \multicolumn{5}{c|}{Lennard-Jones fluid} &
		\multicolumn{5}{c}{polymer melts} \\
		\cline{2-11}
		& \multicolumn{3}{c|}{elapsed time} & \multicolumn{2}{c|}{speedup}& \multicolumn{3}{c|}{elapsed time} & \multicolumn{2}{c}{speedup}\\
		\cline{2-11}
		& \textsc{orig} & \textsc{soa} & \textsc{vec} & \textsc{soa} & \textsc{vec}& \textsc{orig} & \textsc{soa} & \textsc{vec} & \textsc{soa} & \textsc{vec} \\
		\midrule
1 & 288.3 & 169.1 & 68.6 & 1.70 & 4.20 & 179.1 & 109.1 & 77.8 & 1.64 & 2.30 \\
2 & 147.7 & 85.9 & 35.1 & 1.72 & 4.21 & 94.6 & 57.6 & 44.9 & 1.64 & 2.11    \\
4 & 76.8 & 43.7 & 17.8 & 1.76 & 4.32 & 51.0 & 29.1 & 23.4 & 1.75 & 2.18     \\
8 & 40.7 & 22.6 & 9.7 & 1.80 & 4.19 & 27.8 & 15.0 & 12.3 & 1.85 & 2.26      \\
16 & 22.5 & 12.0 & 6.4 & 1.88 & 3.53 & 16.9 & 8.2 & 6.8 & 2.06 & 2.48       \\
32 & 11.4 & 6.3 & 3.1 & 1.79 & 3.71 & 7.7 & 4.1 & 3.5 & 1.89 & 2.22         \\
36 & 10.0 & 5.8 & 3.2 & 1.72 & 3.10 & 6.5 & 3.5 & 3.1 & 1.84 & 2.11         \\
64 & 6.7 & 3.6 & 2.2 & 1.83 & 3.05 & 4.4 & 2.3 & 2.1 & 1.89 & 2.09          \\
72 & 5.9 & 3.4 & 2.1 & 1.71 & 2.83 & 3.8 & 2.1 & 1.9 & 1.82 & 1.97          \\
		\bottomrule
	\end{tabular}
}
\end{table}
\begin{table}[h!]
\caption{Raw data of sectional runtimes to simulation in LJ fluid and polymer melts on \textbf{Raven}.}
\centering
\begin{tabular}{l |ccc|ccc }
	\toprule
	\multirow{2}{*}{Section} & \multicolumn{3}{c|}{Lennard-Jones fluid} &
	\multicolumn{3}{c}{polymer melts} \\
	\cline{2-7}
	& \textsc{orig} & \textsc{soa} & \textsc{vec}& \textsc{orig} & \textsc{soa} & \textsc{vec}\\
	\midrule
Pair & 1.72 & 1.42 & 0.53 & 0.66 & 0.31 & 0.18      \\
FENE & / & / & / & 0.19 & 0.10 & 0.10                \\
Angle & / & / & / & 0.23 & 0.19 & 0.21               \\
Comm & 1.15 & 0.36 & 0.38 & 0.89 & 0.25 & 0.25      \\
Integrate & 0.18 & 0.10 & 0.21 & 0.26 & 0.07 & 0.13 \\
Neigh & 1.86 & 0.60 & 0.13 & 0.45 & 0.17 & 0.06     \\
Resort & 0.45 & 0.79 & 0.73 & 0.43 & 0.80 & 0.80    \\
Other & 0.50 & 0.15 & 0.09 & 0.66 & 0.19 & 0.20     \\
	\bottomrule
\end{tabular}
\end{table}
	\clearpage
	
	\begin{figure}[h!]
		\centering
		a) \includegraphics[width=0.75\textwidth]{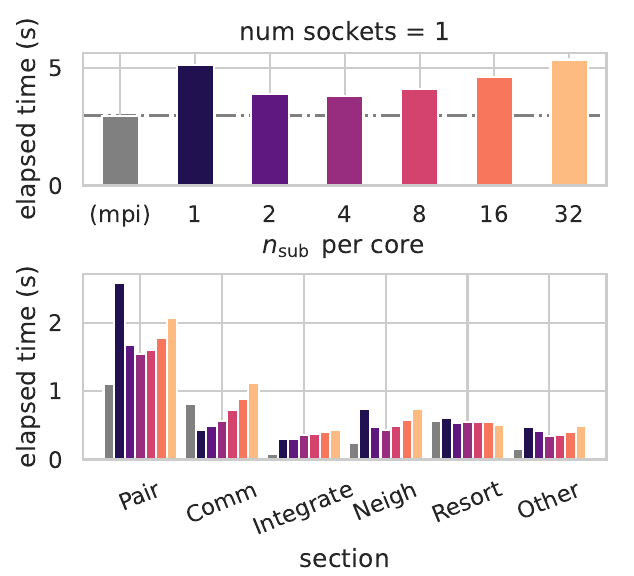}
		\\b) \includegraphics[width=0.75\textwidth]{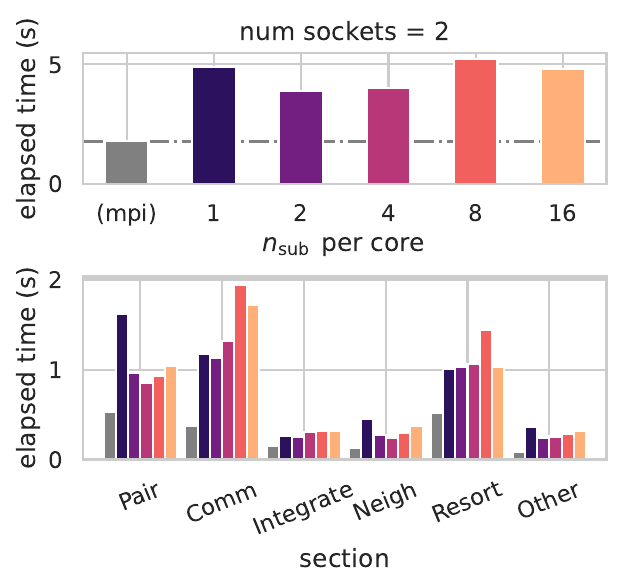}
		
		\caption{Elapsed time for the homogenous Lennard-Jones fluid with the MPI version ({\color{gray}-$\cdot$-$\cdot$}) and with HPX using different number of subnodes on \textbf{Raven}. This figure is supplementary to Fig. 7 in the main text.}
		\label{fig-si:mpi-vs-hpx-bulk-lj-raven}
	\end{figure}
	
	\begin{figure}[t!]
		\centering
		\includegraphics[width=0.9\textwidth]{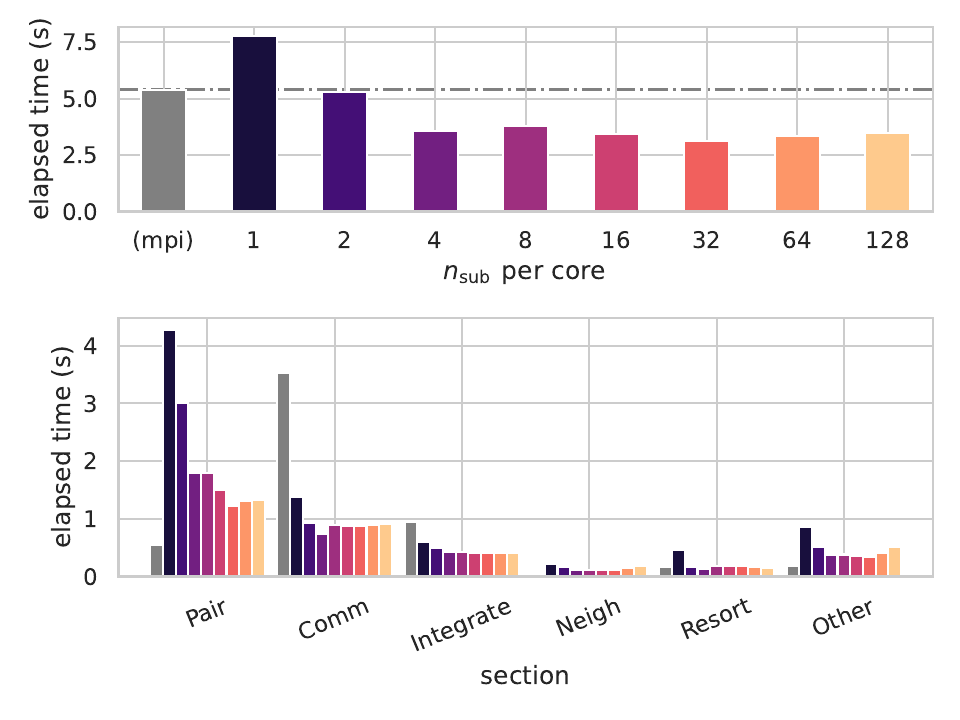}
		\caption{Comparison of sectional elapsed times with the MPI version ({\color{gray}-$\cdot$-$\cdot$}) and with HPX using different number of subnodes for the spherical Lennard-Jones on \textbf{Raven}. This figure is supplementary to Fig. 9 in the main text.}
		\label{fig-si:mpi-vs-hpx-sphere-lj-raven}
	\end{figure}
	\clearpage
	